\documentclass{article}
\usepackage{authblk}
\usepackage{graphicx}%
\usepackage{multirow}%
\usepackage{amsmath,amssymb,amsfonts}%
\usepackage{amsthm}%
\usepackage{mathrsfs}%
\usepackage[title]{appendix}%
\usepackage{xcolor}%
\usepackage{textcomp}%
\usepackage{manyfoot}%
\usepackage{booktabs}%
\usepackage{algorithm}%
\usepackage{algorithmicx}%
\usepackage{algpseudocode}%
\usepackage{listings}%
\usepackage{placeins}
\usepackage{hyperref}

\usepackage{manfnt}
\usepackage{array}

\usepackage[square,numbers]{natbib}

\begin{document}

\title{\Huge Inherent structural descriptors\\ via machine learning}

\author[1]{Emanuele Telari}
\author[1,*]{Antonio Tinti}
\author[1]{Manoj    Settem}
\author[2]{Morgan   Rees}
\author[2,3]{Henry  Hoddinott}
\author[4]{Malcolm  Dearg}
\author[5]{Bernd    von Issendorff}
\author[3]{Georg    Held}
\author[4]{Thomas J.A. Slater}
\author[2]{Richard E.  Palmer}
\author[6,7]{Luca   Maragliano}
\author[8]{Riccardo Ferrando}
\author[1]{Alberto  Giacomello}

\affil[1]{Dipartimento di Ingegneria Meccanica e Aerospaziale, Sapienza Università di Roma, Via Eudossiana 18, Roma, 00184, Italy}

\affil[2]{Nanomaterials lab, Mechanical Engineering, Swansea University, Bay Campus, Fabian way, Swansea, SA1 8EN, UK}

\affil[3]{Diamond Light Source, Harwell Science and Innovation Campus, Fermi Ave, Didcot, OX11 0DE, England}

\affil[4]{Cardiff Catalysis Institute, School of Chemistry, Cardiff University, Translational Research Hub, Cardiff, CF24 4HQ, Wales}

\affil[5]{Department of Physics, Albert-Ludwigs-Universität, Freiburg in Breisgau, 79098, Germany}

\affil[6]{Dipartimento di Scienze della Vita e dell'Ambiente, Università Politecnica delle Marche, Via Brecce Bianche, Ancona, 60131, Italy}

\affil[7]{Center for Synaptic Neuroscience and Technology, Istituto Italiano di Tecnologia, Largo Rosanna Benzi 10, Genova, 16132, Italy}

\affil[8]{Dipartimento di Fisica, Università di Genova, Via Dodecaneso, Genova, 16146, Italy}

\affil[*]{Corresponding author: \texttt{antonio.tinti$\mathtt{\left [at \right ]}$uniroma1.it}}

\maketitle

\begin{abstract}
Finding proper collective variables for complex systems and processes is one of the most challenging tasks in simulations \cite{bolhuis2002}, which limits the interpretation of experimental and simulated data \cite{husic2018markov} and the application of enhanced sampling techniques \cite{sprik1998free,laio2002escaping}. 
Here, we propose a machine learning approach able to distill few, physically relevant variables by associating instantaneous configurations of the system to their corresponding inherent structures as defined in liquids theory \cite{stillinger1983inherent}.
We apply this approach to the challenging case of structural transitions in nanoclusters \cite{wales1998archetypal}, managing to characterize and explore the structural complexity \cite{baletto2005rev} of an experimentally relevant system constituted by 147 gold atoms \cite{telari2023}. Our inherent-structure variables are shown to be effective at computing complex free-energy landscapes, transition rates, and at describing non-equilibrium melting and freezing processes. The effectiveness of this
machine learning  strategy guided by the generally-applicable concept
of inherent structures \cite{stillinger1983inherent} shows promise to devise collective variables for a vast range of systems, including liquids \cite{sciortino1999inherent}, glasses \cite{tanaka2019revealing}, and proteins  \cite{nakagawa2006inherent}.
\end{abstract}


\section{Introduction}
Describing atomic/molecular processes 
is a notoriously difficult endeavour \cite{bolhuis2002} even in apparently simple cases such as 
the isomerization of a small molecule \cite{bolhuis2000reaction}.
Producing a low-dimensional representation of such processes usually requires the introduction of functions of the system coordinates, called collective variables (CVs). CVs can be exploited in advanced simulation techniques \cite{frenkel1984new,laio2002escaping, maragliano2006temperature} for accelerated sampling, FE calculations, and identification of transition mechanisms for a variety of phenomena, including transitions in hard \cite{frenkel1984new}, 
soft and biological \cite{bolhuis2000reaction} matter,
and chemical reactions. 
More generally, starting from the unpractical description in terms of atomic coordinates, CVs attempt to distill essential physical information about complex processes including non-equilibrium ones \cite{allen2009forward}. 

\begin{figure}[h!]
 \centering
 \makebox[\textwidth][c]{\includegraphics[width=18cm]{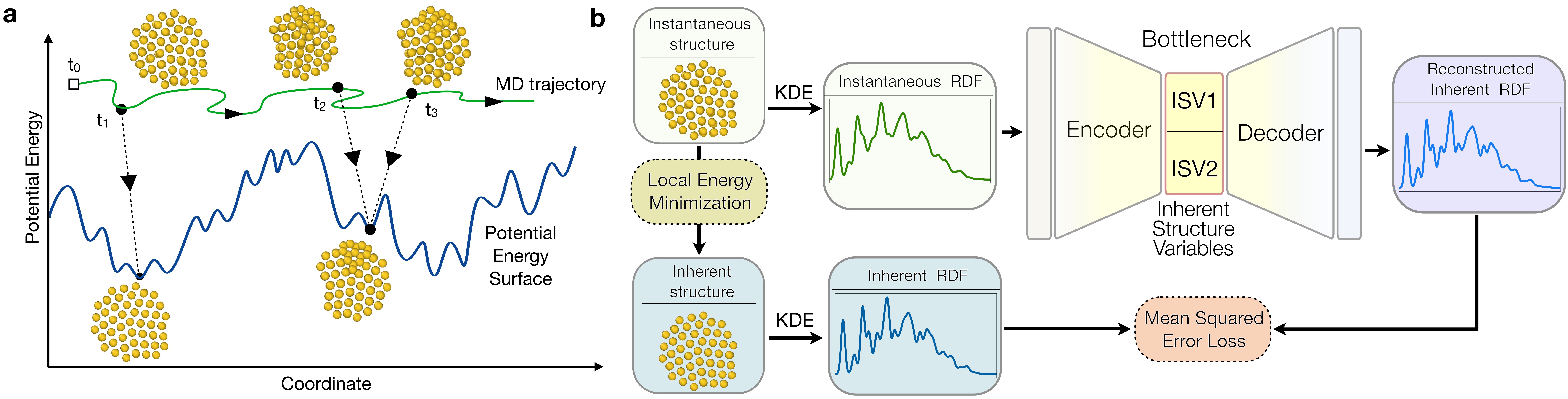}}
 \caption{ \textbf{Proposed approach to distill inherent structural variables.}
 \textbf{a} Schematic representation of the relation between instantaneous configurations at finite temperature and the related inherent structures on the potential energy surface. Instantaneous configurations are affected by significant thermal noise. A local minimization removes the thermal contribution and allows obtaining of the corresponding inherent structure. 
 \textbf{b} Sketch the of working principle behind the ISV encoder-decoder neural network. Structural descriptors of the system, such as the radial distribution function (RDF), constitute the encoder's input, whose output is used by the decoder to reconstruct the inherent state counterpart of the original descriptors. The Mean Square Error (MSE) loss function is used to measure the performance of the network in validation and training.}
 \label{fig:aesketch}
 \end{figure}

Recently, machine learning (ML) has emerged as an invaluable tool for the  discovery of CVs \cite{sultan2018automated,wehmeyer2018time,sidky2020machine,jung2023machine}
in overly complicated systems or when physical intuition fails. In this work, we introduce a generally applicable ML approach for characterizing structural transitions of actual physical systems. We define CVs capable of discriminating structural motifs in noisy finite-temperature configurations based on their zero-temperature counterparts, taking inspiration from the inherent structure concept which we borrow from the theory of liquids \cite{stillinger1983inherent} (Fig.~\ref{fig:aesketch}a). 
In order to devise few, physically informed CVs, we employ a neural network characterized by the  convergent-divergent architecture typical of autoencoders. We train the network such that, while taking  structural descriptors evaluated on finite-temperature realizations as inputs, it learns to associate them to the zero-temperature counterparts of the original descriptors in the output (Fig.~\ref{fig:aesketch}b). The resulting latent variables, which we call inherent structure variables (ISVs), can be thus computed on-the-fly during the dynamical evolution of a system. In addition, they offer a unified description of instantaneous configurations belonging to different temperatures, as they refer to the associated inherent configurations.
For the same reason, ISVs can be adopted to describe both equilibrium and non-equilibrium conditions. ISVs are therefore well suited for phase space exploration, FE and rate calculations or trajectory analysis, and of general interest for any system in which structural diversity is an issue, e.g., nanoclusters, \cite{pavan2015NPsMeetMetaD,tribello2017} bulk crystals \cite{frenkel1984new,pipolo2017navigating}, glasses \cite{fan2017energy,tanaka2019revealing}, and proteins \cite{nakagawa2006inherent,rao2010protein}.

Here, the ISV approach is applied to structural transitions in metal nanoclusters, a challenging class of experimentally-relevant systems characterized by a startling variety of motifs \cite{baletto2005rev}. Indeed, due to their small size, metal nanoclusters can break translational and rotational symmetries, allowing for multiple twinned structures such as icosahedra (Ih) and decahedra (Dh) in addition to standard crystal lattices, as face-centered-cubic (fcc) \cite{baletto2005rev,foster2018AuImagingACFraction}. Moreover, they support several types of surface and internal defects and overall shapes \cite{Mottet1997ss,apra2004AuRosette,Xia2021ncomms}. 

Navigating the structural complexity of clusters has been a long-standing challenge \cite{wales1998archetypal}. 
The fact that dozens of structural families can be identified in metal nanoclusters \cite{telari2023} makes the question about the kinetics and mechanisms of transitions between them even more urgent: how do the atoms of an fcc nanocluster rearrange into a decahedral one? At what rates does such a process take place? Previous studies \cite{pavan2015NPsMeetMetaD,amodeo2020} of structural transitions in metal nanoclusters have relied on carefully identified CVs tailored for a specific transition. However, considering the structural complexity of metal clusters, CVs capable of capturing the fine structural details and navigating the variety of structural motifs is crucial and non-trivial.
Parallel tempering (PT)  
has proven an effective means to explore the structural landscape of coinage metals \cite{settem2022AuPTMD,settemAgCu} exploiting configuration exchanges between replicas at different temperatures to overcome FE barriers without the need of specifying CVs. Building upon this large database of structures, we recently used ML to construct a low dimensional representation of such a landscape that is both physically meaningful and capable of discriminating fine structural details \cite{telari2023}. The key idea was using a translationally and rotationally invariant representation of the cluster, the radial distribution function (RDF), and reduce its dimensionality using a convolutional autoencoder. This approach allowed us to classify locally minimized structures into tens of structural families.

The CVs we defined in \cite{telari2023} are efficient at classifying structures after the removal of thermal noise by means of energy minimization. However, in order to study the finite-temperature evolution of nanoclusters and employ enhanced simulation approaches, CVs able to deal with noisy, finite-temperature configurations are needed. This is the goal of the present work that was achieved by the ML approach introduced above (Fig.~\ref{fig:aesketch}) using RDFs as suitable structural descriptors. 

The  specific system considered in this work is a gold nanocluster of 147 atoms (Au$_{147}$),  which is a magic number for the formation of Ih, Dh, and fcc clusters.  This cluster was specifically selected because it is characterized by the coexistence, over a wide range of temperatures, of a much widervariety of structural motifs with respect to other elemental clusters of similar size \cite{settemAgCu}. This structural wealth was confirmed by scanning transmission electron microscopy (STEM) of size-selected nanoclusters, which is reported below. As a consequence of the competition of many structural motifs, Au$_{147}$ represents a vastly challenging system despite the relatively small number of atoms and the consequent affordability of computer simulations \cite{schebarchov2018AuHSA,settem2022AuPTMD}.

In this paper, we applied our ML approach (Sec.~\ref{sec:ISV}) to compute its FE landscape (Sec.~\ref{sec:US}), rates and mechanisms (Sec.~\ref{sec:rates}) of structural transitions in equilibrium and non-equilibrium (Sec.~\ref{sec:melting}) conditions. 

\section{Inherent structural variables by machine learning}\label{sec:ISV}     

In this section, we devise an approach to build a low-dimensional structural description that enables on-the-fly structural analysis and biasing of molecular simulations. In order to achieve this goal, instantaneous configurations are used as an input, differently from similar approaches aiming at the static classification of Ref.~\cite{telari2023} which rely solely on locally minimized structures. 
The proposed approach for obtaining  descriptors  with a general structural meaning from instantaneous configurations draws inspiration from the inherent structure idea pioneered by Stillinger and Weber for liquids \cite{stillinger1983inherent}: each instantaneous configuration is thought as a fluctuation around the closest local minimum on the potential energy surface \cite{calvo2002}; by quenching, one can refer each dynamical configuration to its \emph{inherent structure} which does not depend on the particular way the original configuration was obtained (equilibrium or non-equilibrium simulations, different temperatures, protocols, etc.), see Fig.~\ref{fig:aesketch}a. 

Similarly to our previous attempts, we leverage RDFs as a convenient translationally and rotationally invariant description of the atomic structures.
From a structural point of view, RDFs are particularly convenient descriptors as they contain extensive information about the structure of clusters \cite{pavan2015NPsMeetMetaD,telari2023} -- both long and short range order -- regardless of whether they are instantaneous configurations or locally minimized ones. The major drawback of RDFs, which has so far limited their use as collective variables, is related to their relatively high dimensionality. This limitation can be alleviated by means of ML.

We propose to couple and compress the high-dimensional information contained in RDFs relative to instantaneous and inherent configurations by means of a conveniently modified convolutional autoencoder. Autoencoders are neural networks with a convergent-divergent architecture as sketched in Fig.~\ref{fig:aesketch}b, well-suited for reducing the dimensionality of data \cite{hinton2006autoencoder}.
In our implementation (Fig.~\ref{fig:aesketch}b), the RDF computed from a specific instantaneous atomic configuration is fed to the autoencoder. In parallel, the same atomic configuration is subjected to a short energy minimization to quench it to the local minimum, resulting in a noise-free RDF. The network is then taught to minimize the mean squared error loss between the output and the inherent-structure RDF, at variance with the usual autoencoder strategy of matching identical inputs and outputs. Thus, in a single step, our network is capable of analyzing instantaneous atomic configurations and match them with their inherent structure, while producing a low dimensional representation. In data science terms, the strategy can be summarized as a classification task where each instantaneous-configuration RDF is labeled according to its inherent counterpart.
In summary, by non-linearly combining information from the input and the output, the  bottleneck obtained by such an approach provides a limited number of descriptors, the ISVs, capable of assigning similar values to different instantaneous configurations  which share the same inherent structure.

As an application, we considered a real-world example, a gold nanocluster consisting of 147 gold atoms, Au$_{147}$. Interactions are modeled via the many-body ``Gupta'' second-moment tight-binding QEq potential \cite{rosato1989}. This potential is known to capture well the of variety structural motifs of gold at this size \cite{telari2023}, which
correspond to those experimentally observed in our STEM data shown in the following. 
We note that other approaches, such as DFT calculations, could in principle be used in conjunction with the  ISVs,  although it is outside the scope of this work which is focused on demonstrating the generality of the approach.
The training set was taken from Ref.~\cite{settem2022AuPTMD}, using Au$_{147}$ configurations generated by parallel tempering at different temperatures. Details about dataset and training are reported in Methods section and Supplementary section Machine Learning. 

The most critical hyperparameter of the autoencoder is the bottleneck size; here we found that the optimal compromise between information compression and reconstruction performance was achieved for a bottleneck of size 2. By comparing the generated space against the structural classification of Ref.~\cite{telari2023}, the two ISVs were found to be expressive enough to encode the fine structural details of the Au nanoclusters (see Supplementary Fig.~1). 

The possibility to compute structural CVs directly from instantaneous structures opens the way to use them  on-the-fly in atomistic simulations, e.g., for analyzing the dynamical structural evolution of the system and to bias trajectories exploiting the intrinsic differentiability of neural networks. 
Results below indeed show that the ISVs are suitable for FE and rate calculations, as well as for analyzing non-equilibrium processes in complex and realistic systems. We were able to handle these diverse applications by training the network only once, as the  description conveniently unifies  information from different temperatures contained in the dataset. 
We further remark that the proposed strategy is general in several ways: 1) it can be used in conjunction with simulations of different kinds, including DFT; 2) due to the flexiblity of neural networks, it can be used on a variety of physical inputs, notably different kinds of spectra; 3) our inherent-structure approach to CVs could prove beneficial for dynamical analysis in other fields,including liquids/glasses\cite{tanaka2019revealing}, and proteins \cite{nakagawa2006inherent}.

\section{Free-energy landscape}\label{sec:US}

\begin{figure}[hb!]
\centering
\makebox[\textwidth][c]{\includegraphics[width=14cm]{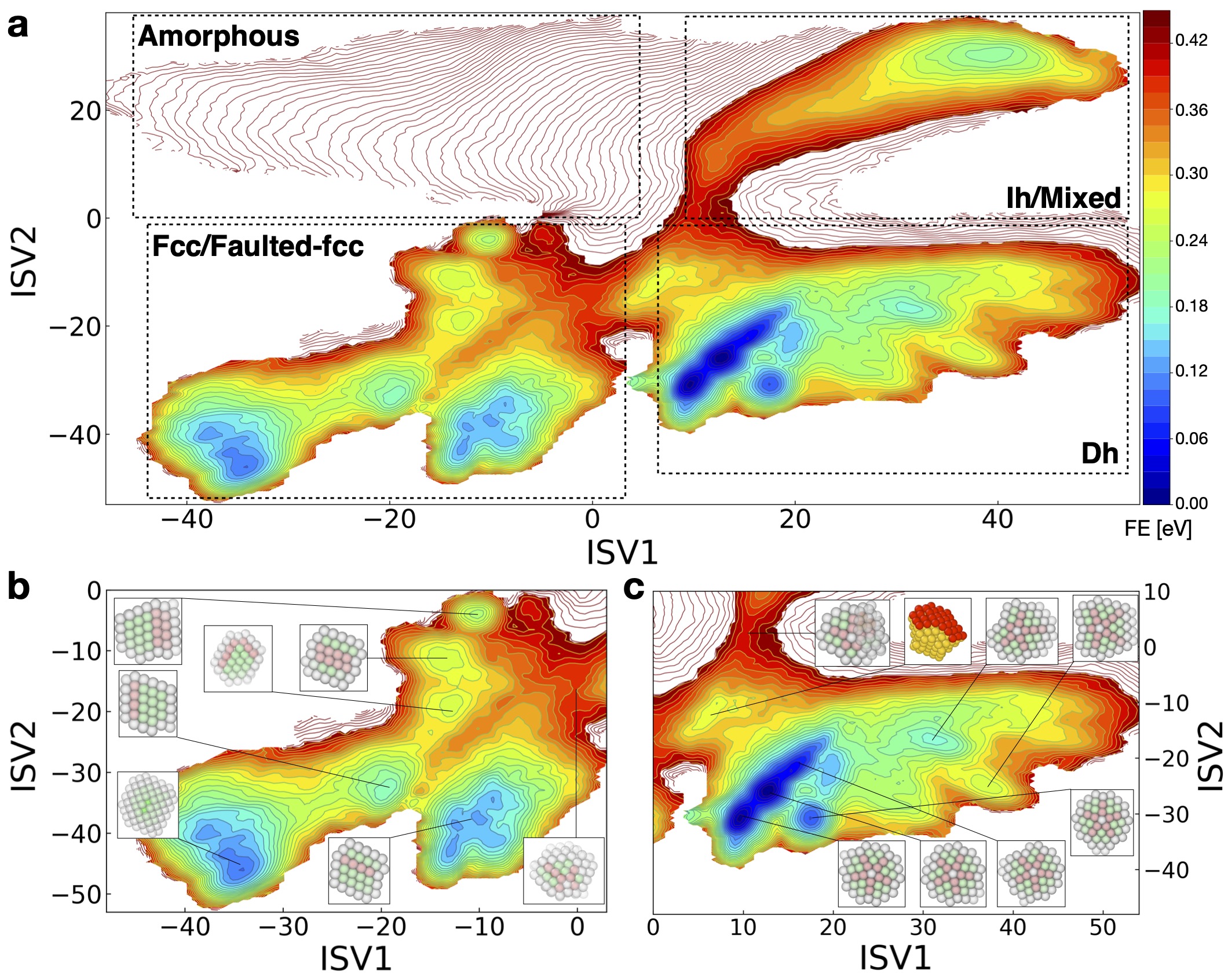}}
\caption{\textbf{FE landscape of Au$_{147}$.}
\textbf{a} Contour plot of the FE landscape obtained by US simulations at 396 K, after WHAM calculations. Color coding of the contours is associated to FE values, as reported in the horizontal colorbar. For values of the free energies above 0.44 eV are displayed only the isolines. The landscape can be divided, according to the FE, in three main regions: fcc and faulted-fcc structures region in the bottom left, Ih and mixed structures region in top right and Dh region in the bottom right. Dh region is connected via a saddle-point to twin and fcc region and via another saddle point to Ih and mixed structure region, while the other two regions are not directly communicating on the landscape. In the top left is where amorphous structures are located, which are associated with very high free energies at this specific temperature. 
\textbf{b} Detailed enlargement of the FE contour plot shown in panel a, showing the fcc and faulted-fcc region together with the representative structures associated with the local minima and the bottleneck linking the fcc region with the Dh basin. 
\textbf{c} Detailed enlargement of the FE contour plot shown in panel a, showing the Dh region and the representative structures associated with the local minima and the bottleneck connecting the Dh region to Ih and Mixed structures. Atoms colored in green, pink, and white have fcc, hcp, and undefined coordination, respectively.}
\label{fig:felandscape}
\end{figure}

We computed the FE landscape of Au$_{147}$ at $400$~K in the 2D space defined by the ISVs obtained by the  strategy illustrated in Fig.~\ref{fig:aesketch}.
We used Monte Carlo Umbrella Sampling simulations in combination with the Weighted Histogram Analysis Method (WHAM) algorithm to unbias the probabilities of approximately $15,000$ restrained simulations spanning all relevant regions of the ISV space. This procedure allowed for the reconstruction the FE landscape reported in Fig.~\ref{fig:felandscape}. Further details are offered in the Methods section and Supplementary Information.  Previous attempts to reconstruct the structural FE landscape of metal nanoclusters were limited to clusters of few atoms \cite{santarossa2010free} or to selected structural transformations \cite{pavan2015NPsMeetMetaD} due to lack of sufficiently informative and low-dimensional CVs capable of comprehensively describing the structural wealth of nanoclusters. 

The FE landscape in Fig.~\ref{fig:felandscape}a offers a high-resolution picture of Au$_{147}$ structures at $0.8$ times the melting temperature. At this temperature, the prevailing structure is Dh, followed by fcc and Ih; amorphous clusters, which occupy the upper left corner, have  large free energies (see the red isolines). As a first approximation, data show that the FE landscape consists of three main basins: fcc, Dh, and Ih. Interestingly, the three basins are connected by two  kinetic bottlenecks (corresponding to the FE saddle points) separating fcc from Dh and Dh from Ih. The Dh basin constitutes a central hub through which all structural transitions at $400$~K are expected to pass. Additionally, although not relevant at this temperature, the mildest slope leading to the amorphous region is found close to the Dh-Ih bottleneck on the Ih side. As we will see, the topology and connectivity between these basins, being based on inherent structural descriptors, offers a general and clear-cut picture of equilibrium and non-equilibrium transitions for Au$_{147}$, including the melting and freezing processes discussed in Section~\ref{sec:melting}.

The kinetic bottleneck separating the fcc basin from Dh is characterized by structures with surface defects. These defected nanoclusters are characterized by the convergence of  two hcp planes (which are the typical feature of twin structures) that give rise to the first seed of a local five-fold axis \cite{Elkoraychy2022nh}, i.e., the distinguishing feature of the decahedral geometry (Fig.~\ref{fig:felandscape}b). 
The saddle point between Dh and Ih features the formation of an hcp island at the surface of an otherwise decahedral cluster (Fig.~\ref{fig:felandscape}c). 

A closer look to the three main basins highlights the presence of multiple local minima in the fcc and Dh basins, which correspond to metastable structures.
The former basin is populated by fcc and various defected structures thereof, chiefly characterized by twinning plane(s)  (Fig.~\ref{fig:felandscape}b). Perfect fcc occupies a rather broad FE minimum at the extreme left. Immediately close to it, a minimum corresponding to a twin cluster with the hcp plane immediately below the surface is found. The basin then forks into a sub-basin on the lower right, which gathers clusters with a single twinning plane in different central positions, and one on the upper right, with multiple minima corresponding to different arrangements of two hcp planes. 

In the Dh basin, the most populated sub-basin corresponds to a central five-fold axis, with multiple local minima pertaining to different kinds of surface defects (Fig.~\ref{fig:felandscape}b). As expected the absolute FE minimum coincides with the perfect Dh structure. Importantly, in between the two main saddle points separating fcc from Dh and Dh from Ih, a local minimum is present characterized by the presence of an hcp island; although characterized by a relatively large FE, this structure occupies a pivotal point for many transitions. Other local minima exist on the far right in which a groove is formed at the cluster's surface.
At this temperature, only one Ih minimum is present, corresponding to the perfect structure. However, a pseudoplateau is present close to the bottleneck, corresponding to mixed structures with mainly amorphous and Ih features \cite{settem2022AuPTMD} that play a major role in melting and freezing (Sec.~\ref{sec:melting}).

\begin{figure}
\centering
\includegraphics[width=1.0\textwidth]{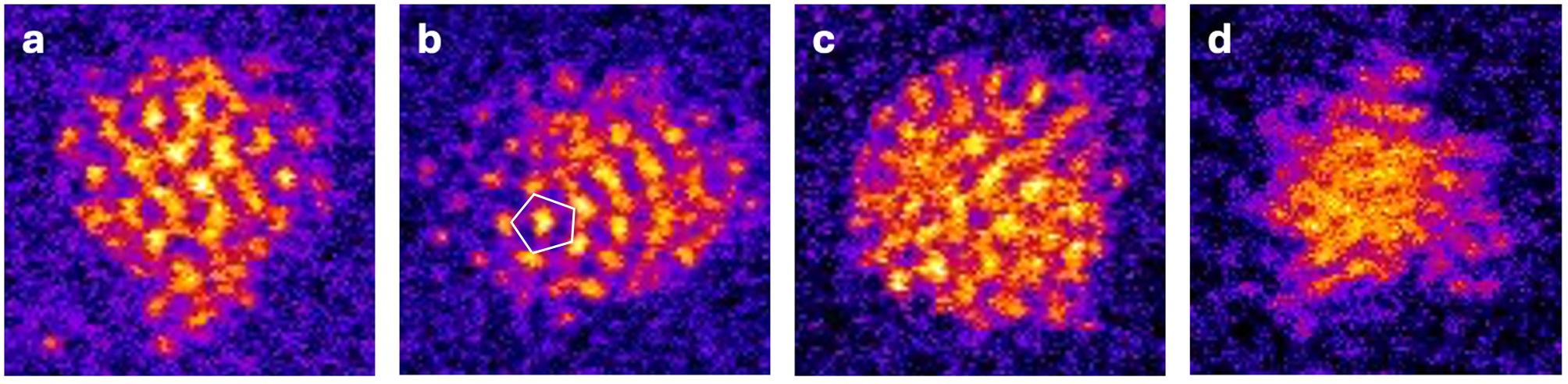}
\caption{
\textbf{Experimental HAADF-STEM micrographs of Au$_{147}$.} Gold nanoclusters are size-selected to be composed of $147\pm 3$ atoms and scans are performed at different temperatures. Representative structures are shown: \textbf{a} fcc (300 $^\circ$C), \textbf{b} decahedron with the local pentagonal arrangement marked (350 $^\circ$C), \textbf{c} icosahedron (300 $^\circ$C), and \textbf{d} amorphous (200 $^\circ$C).}
\label{fig:exp}
\end{figure}

To explore the validity of the simulations, we imaged experimentally size-selected gold clusters, soft-landed onto a carbon support, with the aberration-corrected high-angle annular dark field (HAADF) STEM \cite{wang2012AuImaging}. Imaging at such small sizes poses intrinsic difficulties, due to the fast structural transitions (see next section) and to the interactions of the electron beam with the cluster. 
Nonetheless, results for Au$_{147}$ confirm the main structural families found by simulation: fcc features are clearly visible in several clusters (Fig.~\ref{fig:exp}a); the five-fold axis characteristic of Dh was also detected (Fig.~\ref{fig:exp}b); regular Ih-like structures with arc-like surface features could also be imaged, especially at higher temperatures (Fig.~\ref{fig:exp}c), which is compatible with the melting results presented in Sec.~\ref{sec:melting}. Finally a proportion of amorphous clusters are seen at all temperatures (Fig.~\ref{fig:exp}d).

\FloatBarrier

\section{Transition rates and mechanisms}\label{sec:rates}

\begin{figure}[hbt!]
\centering
\makebox[\textwidth][c]{\includegraphics[width=12cm]{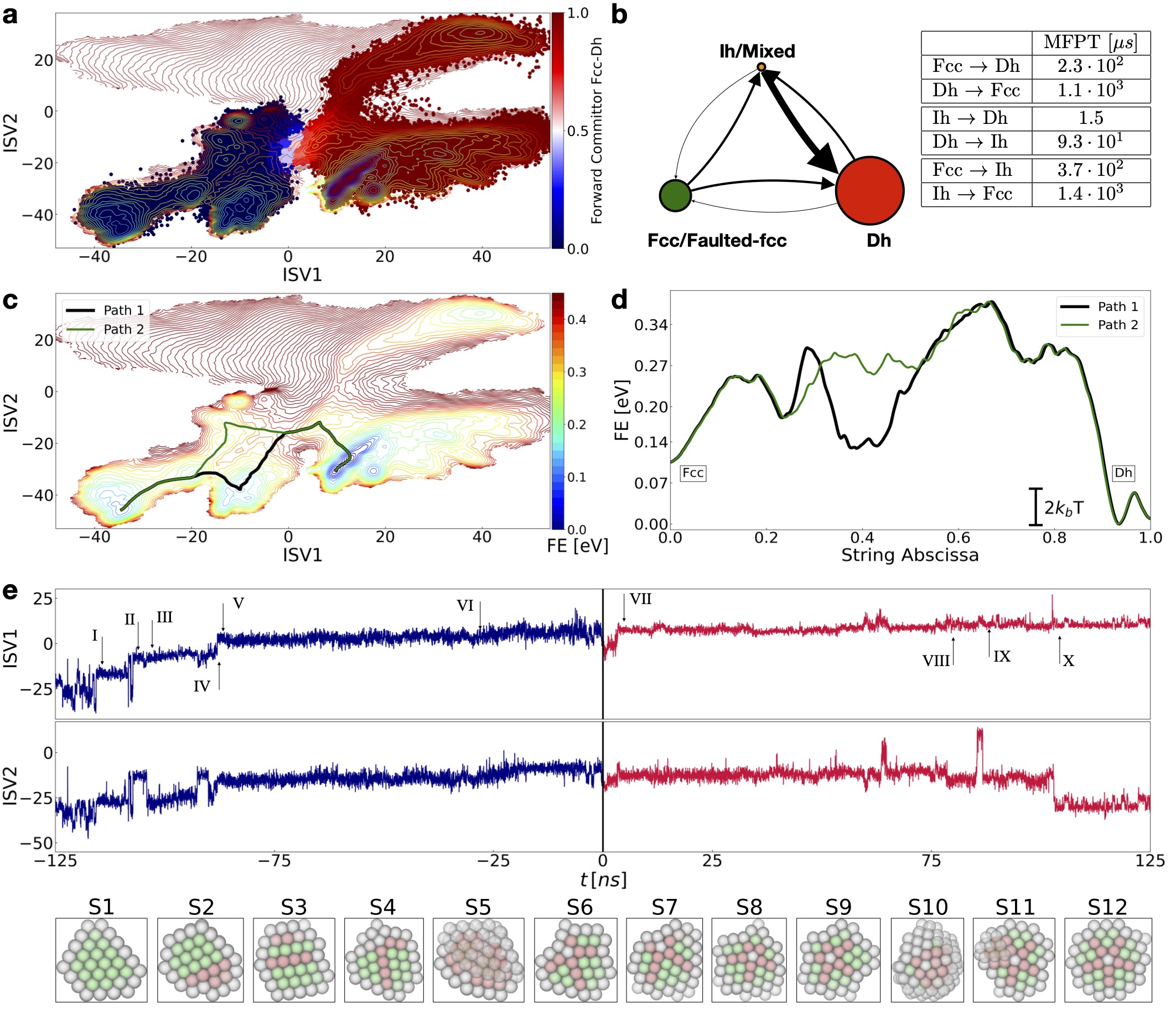}}
\caption{ \textbf{Rates and mechanisms of structural transitions.} 
\textbf{a} Plot of the forward committor $q^{+}$ values of the fcc-Dh transition for different regions of the landscape computed with TPT and MSM. 
\textbf{b} Graph plot of the rates between the three macro region of the ISVs space, representing fcc and faulted-fcc structures (green), Ih and mixed structures (orange) and Dh structures (red). Mean First Passage Times (MFPT) are written in the table. The arrows width is proportional to the log of the ratio between the rates of the transitions (equal to MFPT$^{-1}$) and the smallest rate (Ih$\rightarrow$fcc). The circles areas are proportional to the equilibrium probabilities of finding the system in one of the three states.
\textbf{c} Countour plot with only isolines of the FE where two different optimal paths to go from the fcc basin to Dh absolute minimum are reported. Colorcode of the contour is the same of Fig~\ref{fig:aesketch}a and it is reported in the colorbar on the right.
\textbf{d} FE profile along the two paths shown in panel c. The scalebar reports the conversion to $k_{b}T$ units of the FE.
\textbf{e} Plot versus time of the two ISVs along two different MD unbiased trajectories, initialized in the $q^{+}=0.5$ region. The two trajectories share same initial structure and opposite initial velocities and one is plotted in red for positive times, while the other one is plotted reversed for negative times. Continuous thick black line marks the starting time of both trajectories, while dashed lines marks the most notable transition steps. Atoms colored in green, pink, and white have fcc, hcp, and undefined coordination, respectively.}
\label{fig:rates} 
\end{figure}

The description offered by ISVs, other than performing FE calculations, allows in general to have an on-the-fly description of the dynamics of the system in the low dimensional space. This feature can be exploited to gather information from unbiased trajectories and their dynamical evolution. We made use of that in order to complete the picture offered by the FE landscape and gather information about the kinetics of the main transitions highlighted by the landscape of Fig.~\ref{fig:felandscape}a. We applied very established methods to obtain such information, namely Markov State Models (MSM) \cite{bowman2013introduction} together with Transition Path Theory (TPT) \cite{metznerTPT}. These methods rely on the analysis of a great collection of relatively short unbiased trajectories, described by an appropriate set of observables, in order to quantify the timescales of the slowest processes of a system. Thanks to the ISVs, we were able to launch a wealth of unbiased trajectories, about 4000, distributed over the most relevant regions of the space and track their dynamical evolution, which then has been fed to the aforementioned analysis tools (see Supplementary Figs.~7-10).

Analysis of unbiased trajectories allowed us to compute the committor for the transitions between the three major structural families, i.e. Ih-Dh and fcc-Dh transition (see Supplementary Figs.~11-12). Given two states A and B, simply defined as regions in the CV space, for the $A \to B$ transition the forward committor can be defined as $q^{+}=P(\tau_{B}<\tau_{A})$,  $\tau_{B}$ and $\tau_{A}$ being the time intervals needed for a trajectory to visit basin B or A, respectively. In a similar way, the backward committor can be defined as $q^{-}=P(\tau_{A}<\tau_{B})=1-q^{+}$.  In a nutshell, the forward committor  measures the probability that a trajectory started at a given point in the ISV space ends up in state B ($q=1$) rather than A ($q=0$). For instance, the forward committor for the fcc/Dh transition reported in Fig.~\ref{fig:rates}a shows that trajectories started in the Ih or Dh basins most likely fall in the Dh basin, while those initialized in the fcc one will fall in fcc. While this is intuitive, the most important finding is the exact matching between the region where $q=0.5$ computed from unbiased trajectories and the saddle point in the FE landscape  computed independently by US (the fcc/Dh bottleneck). Similarly, the committor for Ih-Dh transition test has been found to be in good agreement with the FE landscape (Supplementary Fig.~12b). This provides a strong validation of the quality of the ISVs, which describe these processes without the artefacts due to insufficient CVs \cite{bolhuis2002}.

In addition to the committor, overall transition rates and mean first passage times between the three most relevant basins were computed. The rates were estimated by feeding to the MSM the stationary probability distribution computed by means of US simulations (Fig.~\ref{fig:felandscape}).
The  states chosen for this analysis correspond to the three main basins (Dh, fcc, and Ih) which are also those relevant for experiments. The plot in Fig.~\ref{fig:rates}b shows that Ih-Dh is the fastest transition, happening in ca. 1.5~$\mu s$; the related FE barrier is $\Delta F=6\,k_BT$, leading to an estimated prefactor $t_0= 3$~ns, assuming an Arrhenius kinetics for the mean first passage time, $\text{MFPT}=t_0\exp{(\Delta F/k_BT)}$, with $k_B$ the Boltzmann constant and $T$ the absolute temperature. On the other hand, the fcc-Dh transition, which is characterized by a slightly higher barrier $\Delta F=8\,k_BT$, takes more than 100 times more, which corresponds to a much larger effective prefactor, $t_0 = 77 $~ns. This difference can be understood if one considers the presence of multiple local minima in the fcc/twin basin that effectively slows down diffusion to reach the kinetic bottleneck with Dh. 

In Fig.~\ref{fig:rates}c we report the most probable paths joining the fcc minimum with the Dh one, computed by the string method \cite{e2002string} applied on the FE landscape. At least two independent paths are possible, one passing through the simple twin minimum and one through the region corresponding to clusters with multiple hcp planes. Even though both pass through the same transition state, the latter path corresponds to the energetically favored option, as the intermediate barriers are lower (Fig.~\ref{fig:rates}d). While these paths are the statistically most relevant ones in the limit of low thermal noise, the mechanism of the transformation can be observed dynamically and in atomic detail by selecting reactive trajectories for the same transition. This can be done by launching unbiased trajectories from the relevant transition state ($q^{+}=0.5$) with different initial velocities and stitching together two reactive branches ending up in the products ($q^{+}=1$) or in the reactants ($q^{+}=0$). 

Fig.~\ref{fig:rates}e shows two trajectories initialized with opposite initial velocities, sharing the same initial configuration, selected in the transition state region. In such a way, due to the reversibility of the dynamics, the two trajectories can be seen as two portions of the same dynamical evolution single reactive trajectory connecting fcc to Dh, going forward and backward in time.  On the ISV plots, we mark the times corresponding to the key structural changes (indicated by the Roman numerals \emph{I} to \emph{X}) during the transition.

In the initial phase (up to \emph{I}), the cluster fluctuates between fcc (S1) and hcp islands. At this point an increase in ISV1 coincides with the development of peripheral stacking fault (SF) with partial \{111\} surface facet (S2). Very briefly (\emph{II} to \emph{III}), both ISV1 and ISV2 increase resulting in a higher fraction of SF in the cluster and then there is a reduction in ISV2 without much appreciable change in ISV1 (\emph{III} to \emph{IV}) leading to the cluster adopting a twin plus hcp island (S3) and to a lesser extent twin structures. In this duration the cluster makes an excursion to the SF again where ISV2 increases and back. The parallel twin (plus hcp island) develops into a peripheral Dh (S4) around \emph{IV}. A sharp rise in ISV1 around \emph{V} coincides with the increase of the length of some of the twin planes emanating from the Dh axis at the expense of the longest twin plane. Around \emph{VI}, ISV2 increases slightly in correspondance to the formation of hcp islands or an additional 5-fold axis when the hcp islands are on the adjacent \{111\} facets (S5) which persists up to to the transition point at 0 ns. Moving on to the forward branch, around \emph{VII}, we observe the annihilation of an exiting peripheral 5-fold axis (S6) and creation of another peripheral 5-fold axis (S7). A further rearrangement pushes this 5-fold axis inwards (S8). The cluster remains in this arrangement for a long period (up to \emph{VIII}) with appearance and disappearance of hcp islands. A spike in ISV2 around 64 ns results in the cluster adopting a mixed structure (Dh and Ih features co-exist) very briefly. A slight dip in ISV2 at \emph{VIII} results in rearrangement of the surface fcc islands which move away from the Dh axis (S9). This persists up to the beginning of another spike in ISV2 which indicates a transformation into mixed structure (S10) with three 5-fold axes. After this the cluster transforms back (\emph{IX}) into Dh structure with equi-length twins (S11). A final lowering in ISV2 around \emph{X} takes the cluster into best Dh minimum where the cluster adopts the global minimum structure (S12).

{Overall, the above fcc $\rightarrow$ Dh transition can be summarized as $-$ fcc initially forms faulted structures with twins/stacking faults which then leads to the formation of a peripheral 5-fold axis. This undergoes further rearrangement with the 5-fold axis moving inwards and a quick excursion to the Ih/mix region leading to Dh with equi-length twins which eventually rearranges to the global minimum Dh. The initial part of the transition from fcc to peripheral Dh was previously observed \cite{Huang2018prm} in Cu$_{170}$ and Ag$_{146}$ nanoclusters. The twin $\rightarrow$ Dh transition in Au$_{147}$ analyzed using disconnectivity graphs \cite{schebarchov2018AuHSA} suggested that the transition proceeds via disordering with multiple 5-fold axes (i.e. Ih/mix structures). In contrast, our results show that disordering is not necessarily needed to go from twin to Dh. However, we observed that a quick transformation to Ih/mix structures lead to the formation of Dh with equi-length twins which is a feature of the global minimum.}

\FloatBarrier

\section*{Venturing into non-equilibrium: melting and freezing}\label{sec:melting}

\begin{figure}[h!]
\centering
\makebox[\textwidth][c]{\includegraphics[width= 0.8\textwidth]{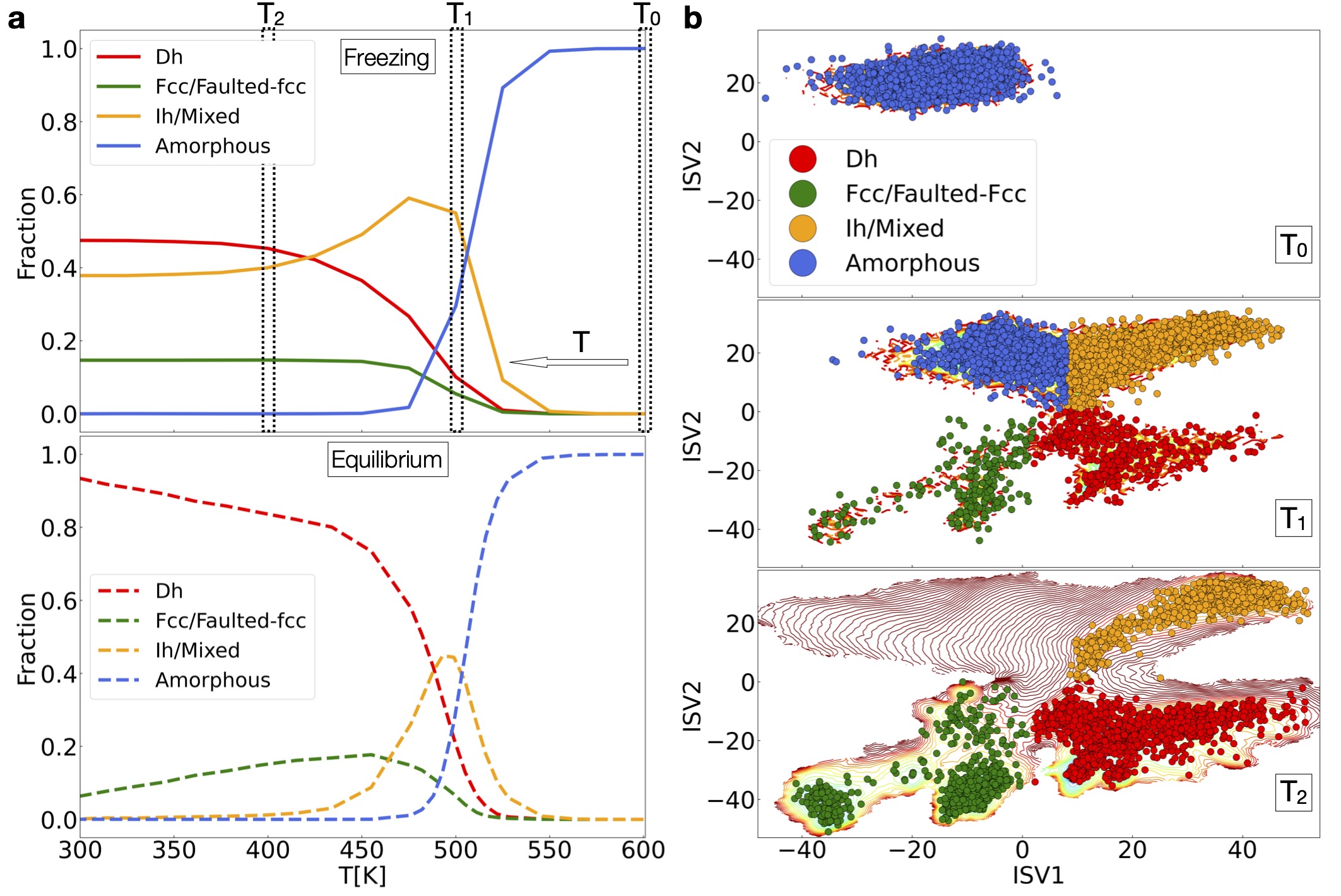}}
\caption{ \textbf{Freezing of Au$_{147}$.}
\textbf{a} Comparison between fractions of main structural families vs. temperature observed in freezing simulations (top panel) and equilibrium simulations (bottom panel) performed in the work of Ref.~\cite{settem2022AuPTMD}.
The structures have been split in the same 4 structural families shown in Fig.~\ref{fig:felandscape}a. Fractions of amorphous structures are reported by the blue lines, fcc/faulted-fcc by green lines, Dh by red lines and Ih by orange lines. 
\textbf{b} Instantaneous distributions for all freezing trajectories at three specific temperatures, T$_{0}=600$ K, T$_{1}=500$ K, T$_{2}=400$ K, highlighted also in the non-equilibrium fractions plot of panel a. For the plots of T$_{0}$ and T$_{1}$ below the points there are the contour plots of the log densities of PT-MD data from Ref.~\cite{settem2022AuPTMD} at those specific temperatures. For the plot of T$_{2}$ the contour plot of the FE shown in Fig.~\ref{fig:felandscape}a is shown. Points are colored according to their structure type using the same color code as in  panel a.}
\label{fig:noneq_freeze} 
\end{figure}

ISVs obtained exploiting the inherent structure description are also feasible to analyze and drive non-equilibrium simulations. To demonstrate this point, we performed freezing and melting simulations for Au$_{147}$, imposing an increasing or decreasing, respectively, temperature ramp of $1$~K/ns to thousands of independent replicas, see the Methods section for details. Figure~\ref{fig:noneq_freeze} shows the distribution of structures observed at different temperatures during the freezing process. 

A merit of the ISV space is to allow the visualization of the time-evolution of structural populations, which can then be compared against the corresponding equilibrium estimates. These equilibrium distributions were taken from US simulations at $400$~K and from the parallel tempering data of Ref.~\cite{settem2022AuPTMD} at other temperatures.  The system starts in the amorphous basin at $600$~K and explores a region which coincides with equilibrium expectations. At lower temperatures ($500$~K), the Ih basin, which is the closest one to amorphous, becomes densely populated, with a prevalence of mixed structures; few trajectories also fall in the Dh and fcc basins. At even lower temperatures ($400$~K), the amorphous population has disappeared, leaving room to the three main structural motifs. Interestingly, when one compares the populations obtained in non-equilibrium and in equilibrium, there is a striking difference concerning Ih and Dh (Fig.~\ref{fig:noneq_freeze}a): Ih are kinetically trapped in the freezing simulations accounting for ca.~$40\%$ of the population, while the equilibrium fraction would be negligible. This happens mainly at the expense of the Dh population that decreases from $80\%$ down to $50\%$ in non-equilibrium (at $400$~K). If the cooling is sufficiently fast (and the temperature is then kept low), it is possible to select Ih clusters. More generally, ISVs produce an intuitive map that could be useful for designing controlled freezing protocols capable of selecting specific polymorphs \cite{amodeo2020} in clusters of different metals and sizes; actually, this approach is expected to be more effective for larger clusters for which the typical transition rates are slower \cite{dearg2024frame}.

\begin{figure}[h!]
    \centering
    \makebox[\textwidth][c]{\includegraphics[width=6cm]{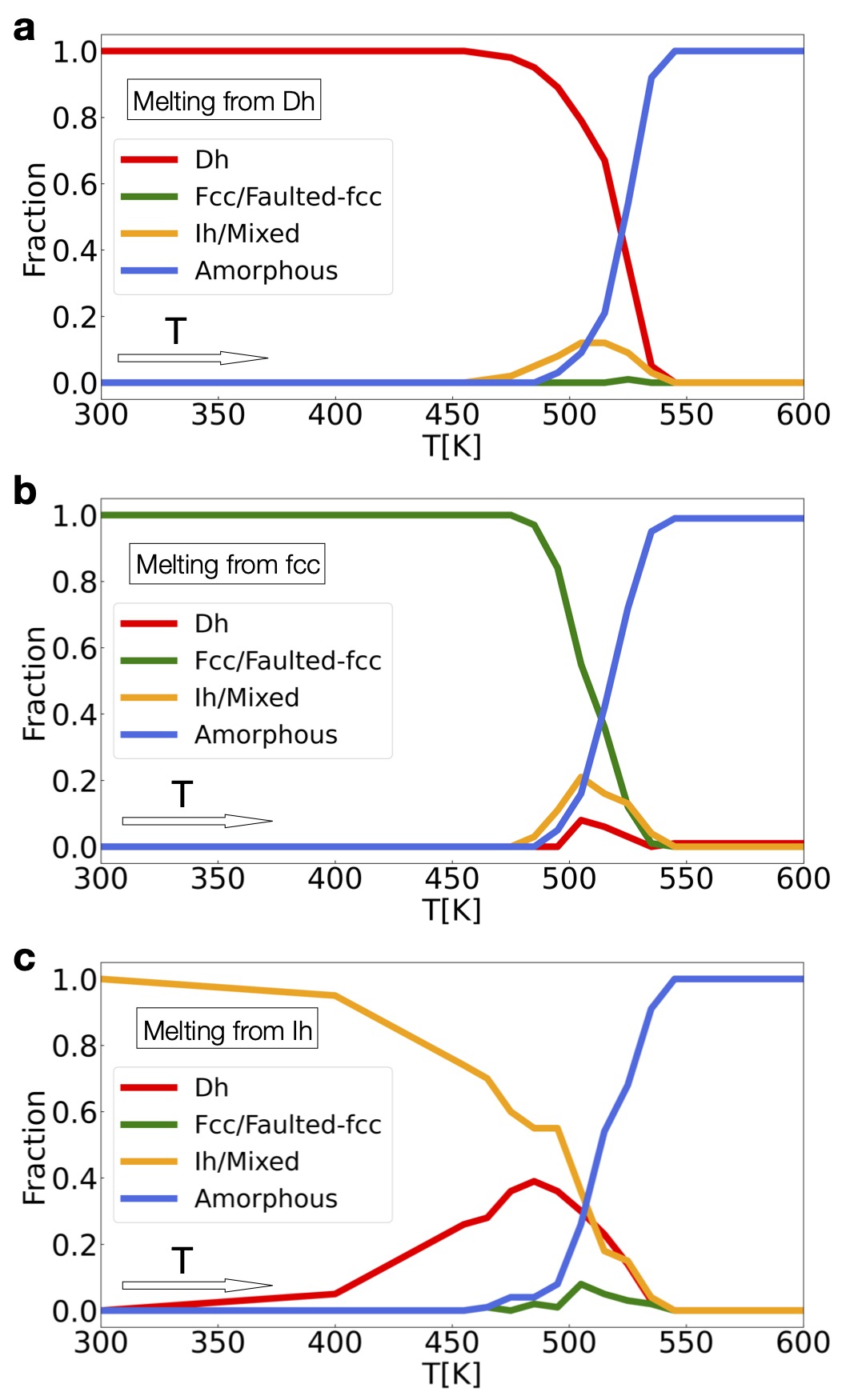}}
    \caption{ \textbf{Melting of Au$_{147}$.}
    \textbf{a} Fractions of main structural families vs. temperature observed in melting simulations initialized from a Dh structure.
    The structures have been split in the same 4 structural families shown in Fig.~\ref{fig:felandscape}a. Fractions of amorphous structures are reported by the blue lines, fcc/faulted-fcc by green lines, Dh by red lines and Ih by orange lines. 
    \textbf{b} Fractions of main structural families vs. temperature observed in melting simulations initialized from an fcc structure.
    Same color code of panel a.
    \textbf{c} Fractions of main structural families vs. temperature observed in melting simulations initialized from an Ih structure.
    Same color code of panel a.
    }
    \label{fig:noneq_melt} 
\end{figure} 

The melting simulations follow a similar protocol, with the initial configurations being extracted from the three main (meta)stable basins. 
To achieve melting of Dh and fcc clusters (Fig.~\ref{fig:noneq_melt}a and 
b), the system has to traverse the mixed region in the Ih basin  which, due to its position, plays a major role in both melting and freezing. Interestingly, if the system is initialized close to a perfect Ih (Fig.~\ref{fig:noneq_melt}c), it still has to traverse the same mixed region but it reaches it for the first time at much lower temperatures. The system is thus able to overcome the Ih-Dh barrier and to populate the Dh basin; Dh then melts with the usual mechanism at higher temperatures ($>500$~K). The crucial role played by mixed/Ih structures close to the melting/freezing temperature is further supported by the fact that they can be observed by HAADF-STEM at high temperatures (Fig.~\ref{fig:exp}c).

\FloatBarrier

\section{Conclusions}\label{sec:Disc}

In conclusion, we devised a ML approach to obtain few general and yet informative collective variables that enable the dynamical analysis of structural transitions. We coupled information from instantaneous atomic configurations and the related inherent structures using autoencoders. This approach distilled a small set of inherent structural variables capable of finely describing the structural landscape, evolution, and transitions of metal nanoclusters, both in equilibrium and in non-equilibrium. Our ISVs, in conjunction with umbrella sampling, 
allowed us to compute a high-resolution two-dimensional FE landscape of Au$_{147}$ nanoclusters, revealing that the topology of the structural space comprises three major FE basins: fcc, Dh, and Ih. Scanning transmission electron microscopy experiments confirmed  the existence of these structures in Au$_{147}$. In addition, our simulations shown that the minima are connected by two kinetic bottlenecks with Dh at the center. The basins are populated by several local FE minima, accounting, at finite temperature, for a wealth of metastable states and structural transitions among them. 
Finally transition rates between the three main FE basins were computed by means of Markov state models, which allowed to validate quality of the ISVs  by means of committor evaluations.
In addition, the two ISVs were capable of tracking the structural evolution of non-equilibrium melting and freezing simulations, rationalizing routes to polymorph selection and recurring melting patterns.
The generality of the strategy supported by the excellent results achieved for metal nanoclusters suggest that ISVs could be used also in different contexts, including the field where the idea of inherent structures originated, i.e., liquids \cite{stillinger1983inherent} but also glasses \cite{tanaka2019revealing}, colloids \cite{de2015entropy}, and proteins \cite{nakagawa2006inherent}.

\section{Methods}\label{sec:Meth}
\subsection*{Deep Learning}
ISVs are obtained by training a modified autoencoder neural network, which
associates to the RDF of a non-minimized structure its inherent counterpart.
The network has the typical convergent-divergent architecture of autoencoders, where the first half, i.e. the encoder, is composed of convolutional layers, while the second half, i.e. the decoder mirrors the encoder and is composed of deconvolutional layers.
The ISVs are obtained as output of the trained encoder. The dataset for the network training is a collection of $613,872$ Au$_{147}$ structures generated by means of Parallel Tempering. Every structure has then been  minimized leading to the computation of the RDF for both instantaneous and inherent structures.

The network has then been trained with a bottleneck size equal to 2, feeding it the non-minimized structures RDFs and comparing the outputs with the associated inherent structures RDFs via a Mean Square Error loss function.

For a more detailed description of the dataset, network architecture and training see the Machine Learning section of the Supplementary Information .

\subsection*{Umbrella Sampling Monte Carlo simulations}

Umbrella sampling simulations were performed using a Metropolis Monte Carlo code which was custom written in C++ for this purpose (see Supplementary Information, Sec. Monte Carlo Code). A total of $15,022$ simulations, distributed all over the ISV landscape have been performed. Simulations have been initialized using the thermalized structure in the training dataset that is closest to the restraining value. After careful tuning the harmonic spring constant of the umbrellas has been set equal to 0.1 eV. The simulations consisted in a total of $20\cdot10^{6}$ MC moves. CVs values have been sampled every $5\cdot10^{3}$ moves for a total of $4\cdot10^{3}$ samples for every simulation.
After discarding the first $1/4$ of samples, random sampling with replacement was used to generate 10 different samplings from the original populations. 
FE was then reconstructed using each of these samplings allowing for the   statistical error estimation via the boostrapping method.
For the reconstruction of the FE landscape for each boostrap realization the  WHAM algorithm \cite{ferrenberg1989} has been used in the implementation by Grossfield \cite{grosswham}. 
A more detailed description of the simulations procedure with information on the convergence of the FE reconstruction is provided in the Supplementary Information section Umbrella Sampling.

\subsection*{Molecular Dynamics and Markov State Models}

MD simulations were performed using the LAMMPS code \cite{LAMMPS}
 augmented with custom Python code to perform on-the-fly estimation of the ISVs.
Simulations have been thermalized using a Langevin thermostat
with a time constant of 1.0 fs.  A $5$ fs timestep was used  during integration. Each integration was carried for  $0.5$ $\mu$s (corresponding to $100\cdot10^{6}$ timesteps), during which the sampling of the ISVs was performed every $50$ ps resulting in a total of $10^{4}$ samples for every simulation. A grand total of $4,448$ of these simulations have been performed, starting from initial configurations distributed all over the most relevant regions of the FE landscape (see Supplementary Fig.~7a). These simulations amounted to a total sampling time of $2.2$ ms. Again simulations have been initialized  by picking the thermalized structure of the training dataset closest to the selected starting point.
MSM\cite{bowman2013introduction} calculations have been performed using the DeepTime \cite{hoffmann2021deeptime} library.
This analysis has been conducted in the ISV space, leveraging information on the stationary distribution obtained by the US calculations.
Committor was estimated using Transition Path Theory\cite{noeCommittor,metznerTPT} as implemented in DeepTime library\cite{hoffmann2021deeptime}.
Additional information regarding MSM is reported in the Supplementary Information section Markov State Models.

\subsection*{Non Equilibrium simulations}
Non Equilibrium MD simulations (freezing and melting) were performed using the LAMMPS code using a Langevin thermostat with the same time settings described in the previous section. The freezing simulations start from a highly disordered liquid configuration which is equilibrated at 600 K for 1 ns. The temperature is then decreased at a rate of 1 K/ns to a final temperature of 300 K. In the case of melting we considered four different initial configurations -- fcc, twin, Ih, and Dh. The initial configurations are equilibrated at 300 K for 1 ns and then the temperature is raised to 600 K at a rate of 1 K/ns. In both the cases, configurations were sampled every 5 ps in the temperature range of 450 K to 550 K and 50 ps at other temperatures. A total of 4200 freezing simulations and 300 melting simulations were performed.

\subsection*{Experimental Methodology}
The Swansea University Nanocluster Source (SUNS), located at the B07 beamline of the Diamond Light Source synchrotron, was used to produce and deposit gold clusters for experimental Scanning Transmission Electron Microscope (STEM) imaging and thus structure comparison with the theoretical results via the Simulation Atlas approach \cite{wang2012AuImaging,wang2012Au20,wang2012Au55}. Size-selected Au147 clusters (N=$147\pm 3$ atoms) were deposited onto silicon nitride heating chips (DENS Solutions) using this DC magnetron-sputtering, inert-gas condensation cluster beam source coupled with a lateral time-of-flight mass selector and deposition stage \cite{pratontep2005,issendorff1999}. The mass filter (resolution M/$\Delta$M=25 ) was calibrated with a beam of Ar+ ions. To reduce cluster agglomeration, the cluster beam was rastered across the support to deposit a uniform coverage (approximately 1\% by projected surface area) on the Silicon Nitride imaging window. Clusters were soft-landed \cite{diVece2005} at a kinetic energy of 1 eV/atom and allowed to diffuse and immobilise at pre-formed defect sites created in advance by sputtering of the window with an Ar+ beam at 500 V for 10 minutes \cite{claeyssens2006}. The agglomeration observed could be associated with harmonics of the incidence Au147 clusters (see below).

HAADF-STEM images were acquired with a JEOL ARM300F (GRAND-ARM) microscope at the electron Physical Sciences Imaging Centre (ePSIC) at Diamond Light Source. The electron beam energy was 300 kV and beam current was approximately 30 pA. The probe semi-angle was approximately 23 mrad and the HAADF detector had an inner collection angle of approximately 58 mrad (outer angle approximately 215 mrad). A DENS Solutions Wildfire holder was used to heat the samples to a range of temperatures (100 $^\circ$C, 150 $^\circ$C, 200 $^\circ$C, 250 $^\circ$C, 300 $^\circ$C and 350 $^\circ$C consecutively). Temperature is monitored by a 4-point probe and is typically stable to within $\pm$1 $^\circ$C; all samples were measured within a central window to ensure accuracy. Videos were acquired using a plug-in for Digital Micrograph, with a frame acquisition time of 1.31 s.

The cluster structure typically fluctuates from frame to frame. The structural assignment of each frame in each cluster video was accomplished by comparison with a Simulation Atlas generated using the abTEM Python package \cite{madsen2021abTEM}. The PRISM algorithm \cite{ophus2017} was used to simulate images (electron energy of 200 keV), a convergence semi-angle of 28 mrad, an interpolation factor of 4 and 10 frozen phonon iterations. Poisson noise was added to the simulated data to approximate an electron fluence of $1\cdot10^{5}$ e$^-$/\AA$^{-2}$ (which is on the order of the fluence used in our imaging). We note that the electron energy and convergence semi-angle of the simulations do not exactly match those of the experiment, but the relevant structural elements used to assign the cluster structures do not depend on the microscope parameters and so the isomers can be assigned regardless.

The Au$_{147}$ clusters were identified as the smallest clusters in each video frame; also found on the surface were larger clusters $-$  being multiples of 147 atoms, as judged by their integrated intensities \cite{wang2011}, presumably formed by surface agglomeration. The illustrative example images shown in Fig.~\ref{fig:exp} are low-noise Au$_{147}$ clusters. The images shown were chosen to illustrate the principal structural motifs observed in the experiments. These images were processed by application of a high frequency filter to suppress noise and adjustment of brightness and contrast. A colour gradient was also mapped onto the greyscale images to better highlight the structural features. The processed frames are compared with the best fits in the simulation atlases for icosahedral, decahedral and fcc structures of an Au$_{147}$ cluster. The atlases cover the full range of polar and azimuthal orientations. Recent examples of this approach are Refs. \cite{foster2018AuImagingACFraction}, \cite{dearg2024} and \cite{lethbridge2024}. Key  to the manual best matching process are the patterns and symmetries in the core region of the nanoparticle, where the signal is highest.

\section{Acknowledgements}
We thank John Russo for useful discussions.

We acknowledge financial support under the National Recovery and Resilience Plan (NRRP), Mission 4, Component 2, Investment 1.1, Call for tender No. 104 published on 2.2.2022 by the Italian Ministry of University and Research (MUR), funded by the European Union – NextGenerationEU– Project Title PINENUT – CUP D53D23002340006 - Grant Assignment Decree No. 957 adopted on 30/06/2023 by the Italian Ministry of University and Research (MUR). We thank Diamond Light Source for access to and support in use of the electron Physical Science Imaging Centre (Instrument E02, Proposal Number: MG28449), and gratefully acknowledge EPSRC grant EP/V029797/2 for support of the electron microscopy.

\section{Author Contributions}
E.T. and A.T. designed the Machine Learning algorithms and programmed the MD/MC simulation codes.
E.T. coded the Neural Networks, performed equilibrium simulations and analyzed simulation data.
A.T. supervised simulation work and data analysis.
M.S. performed structural analysis and non-equilibrium simulations.
E.T., A.T., R.F., and A.G. conceptualized the work.
\vspace{2pt}

\noindent
H.H. produced the clusters under the supervision of B.v.I., G.H. and R.E.P.
M.D. performed ac-STEM imaging of clusters under the supervision of T.S.; 
T.S. produced the Simulation Atlas. M.R. matched experimental images to the simulation atlas under the supervision of R.E.P.;
H.H, B.v.I., G.H., T.S., M.R., and R.E.P. wrote the experimental section of the manuscript under the coordination of R.E.P. .
\vspace{2pt}

\noindent
L.M. supervised rare event methodology,
R.F. and A.G. supervised the work.
A.G. wrote the original draft, which was reviewed and approved by all authors.

\section*{Declarations}
The authors declare no competing interests.

\begin{appendices}
\end{appendices}

\FloatBarrier
\clearpage
\renewcommand{\thefigure}{\arabic{figure}}
\renewcommand{\thesection}{}
\renewcommand{\thesubsection}{S\arabic{subsection}}
\renewcommand{\figurename}{Supplementary Fig.}
\renewcommand{\tablename}{Supplementary Table}
\setcounter{equation}{0}
\setcounter{table}{0}
\setcounter{figure}{0}
\setcounter{section}{0}
\appendix
\section*{Supplementary Information}

\subsection*{Machine Learning}
\label{si:ae}
\subsubsection*{Training Dataset}
The dataset is the same used in our previous work \cite{settem2022AuPTMD}, generated using PTMD with 24 replicas at 24 different temperatures (300, 314, 329, 345, 361, 378, 396, 415, 434,  455, 476, 482, 488, 493, 499, 505, 511, 516, 522, 528, 546, 564, 582, 600 K).
For every replica, 25578 structures were collected, for a total of 613872 structures. Below is reported the number of structures of the dataset divided according to their CNA classification:
\begin{lstlisting}
PTMD data

Dh configurations = 278,405 
Ih configurations = 28,911 
Twin configurations = 29,553 
Fcc configurations = 19,248 
Mix configurations = 69,641 
Amorphous configurations = 188,114 

Total = 613,872
\end{lstlisting}

Every structures was energy minimized, and for every couple of minimized and non-minimized structures the RDF was computed, performing KDE \cite{rosenblattKDE,parzenKDE}, over the interatomic distances using gaussian kernels with bandwidth equal to 0.2.

\subsubsection*{AE architecture}
The network was built and trained using PyTorch library\cite{paszke2017automatic}.

The autoencoder is composed by two main blocks, the encoder and the decoder composed by 5 convolutional layers, and a central block composed by fully connected layers. Input and output layer share the same structures and are two convolutional layers.   

\begin{lstlisting}[basicstyle=\small]
----------------------------------------------------------------
        Layer (type)               Output Shape         Param #
================================================================
            Conv1d-1             [-1, 128, 340]           2,688
         MaxPool1d-2             [-1, 128, 170]               0
              ReLU-3             [-1, 128, 170]               0
       BatchNorm1d-4             [-1, 128, 170]             256
            Conv1d-5              [-1, 64, 170]         122,944
         MaxPool1d-6               [-1, 64, 85]               0
              ReLU-7               [-1, 64, 85]               0
       BatchNorm1d-8               [-1, 64, 85]             128
            Conv1d-9               [-1, 32, 85]          20,512
        MaxPool1d-10               [-1, 32, 42]               0
             ReLU-11               [-1, 32, 42]               0
      BatchNorm1d-12               [-1, 32, 42]              64
           Conv1d-13               [-1, 16, 42]           2,576
        MaxPool1d-14               [-1, 16, 21]               0
             ReLU-15               [-1, 16, 21]               0
      BatchNorm1d-16               [-1, 16, 21]              32
          Flatten-17                  [-1, 336]               0
           Linear-18                    [-1, 2]             674
           Linear-19                  [-1, 336]           1,008
             ReLU-20                  [-1, 336]               0
         Upsample-21               [-1, 16, 42]               0
  ConvTranspose1d-22               [-1, 32, 42]           2,592
             ReLU-23               [-1, 32, 42]               0
      BatchNorm1d-24               [-1, 32, 42]              64
         Upsample-25               [-1, 32, 84]               0
  ConvTranspose1d-26               [-1, 64, 85]          20,544
             ReLU-27               [-1, 64, 85]               0
      BatchNorm1d-28               [-1, 64, 85]             128
         Upsample-29              [-1, 64, 170]               0
  ConvTranspose1d-30             [-1, 128, 170]         123,008
             ReLU-31             [-1, 128, 170]               0
      BatchNorm1d-32             [-1, 128, 170]             256
         Upsample-33             [-1, 128, 340]               0
           Conv1d-34               [-1, 1, 340]           2,561
================================================================
Total params: 300,035
Trainable params: 300,035
Non-trainable params: 0
----------------------------------------------------------------
Input size (MB): 0.00
Forward/backward pass size (MB): 2.21
Params size (MB): 1.14
Estimated Total Size (MB): 3.35
----------------------------------------------------------------
\end{lstlisting}
\FloatBarrier

\subsubsection*{Training}
The dataset was split in training and validation set, using the 20$\%$ of the data as validation set. Then, the dataset was divided in batches of size equal to 128 data. 
The autoencoder was trained using MSE loss function and Adam optimizer \cite{adam}. The starting learning rate was set to 0.004 and then updated using a step scheduler halving its value at epoch number 70 and 90.

\subsubsection*{2D space of training dataset}
\begin{figure}
        \centering
        \includegraphics[width= 12cm]{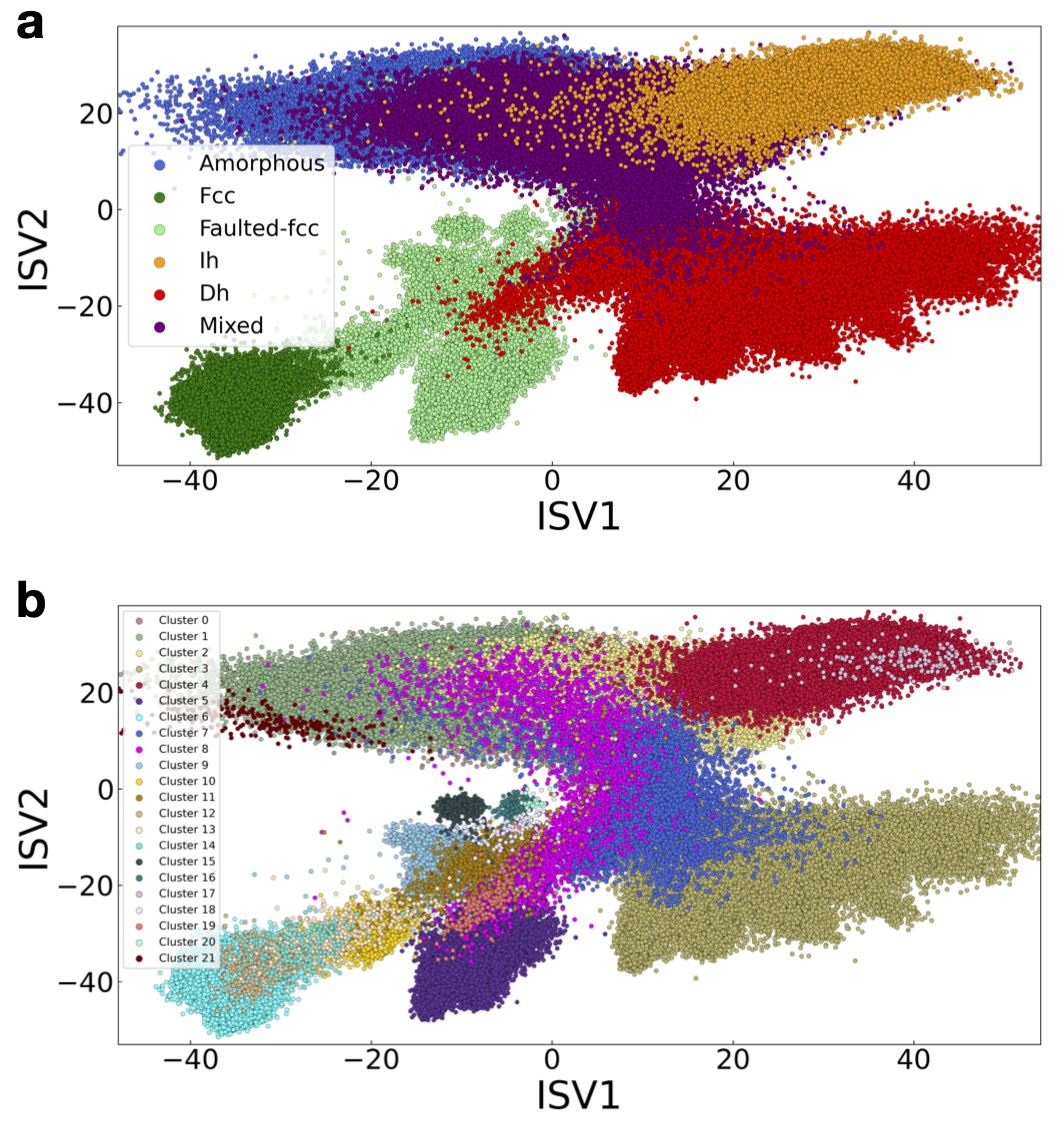}
        \caption[]{\textbf{2D space generated after the training.}
        \textbf{a} PTMD data plotted in the ISV space generated after the training by the AE, colored according to their CNA classification. \textbf{b} PTMD data plotted in the ISV space generated after the training by the AE, colored according to the classification of Au$_{147}$ performed in Telari et al. (2023)\cite{telari2023}. }
        \label{fig:train_embed}
\end{figure}

In Fig.~\ref{fig:train_embed} is reported of the overall PTMD dataset in the ISV space generated after the training.

\FloatBarrier

\subsection*{Umbrella Sampling}
\label{si:us}

\begin{figure}
        \centering
        \includegraphics[width= 10cm]{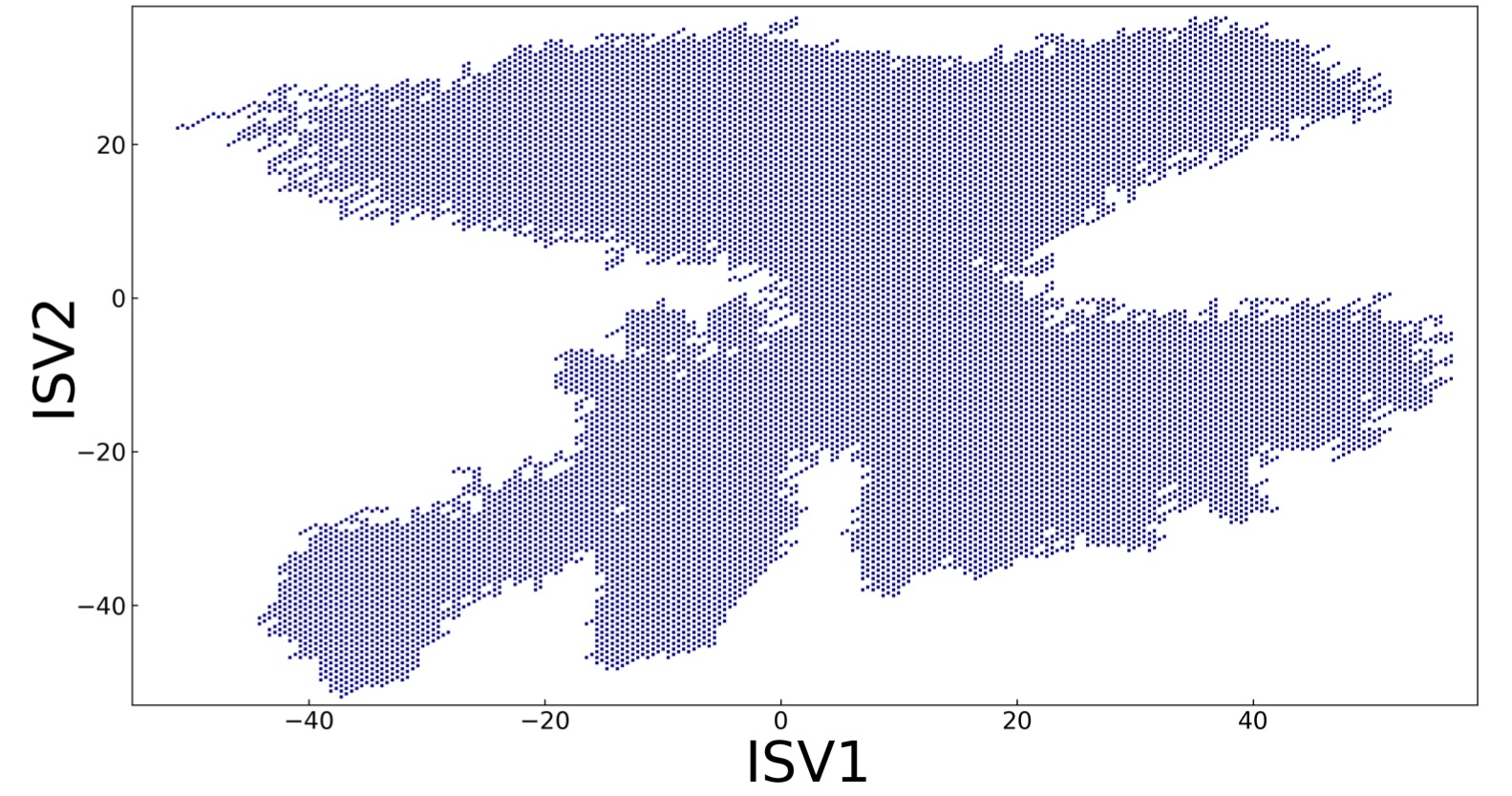}
        \caption[]{\textbf{Grid of the restrained simulations for US.}
        Points in the ISV space where US simulations were  restrained. Every point stands for one restrained simulation.}
        \label{fig:us_points}
\end{figure}

\begin{figure}
        \centering
        \includegraphics[width= 10cm]{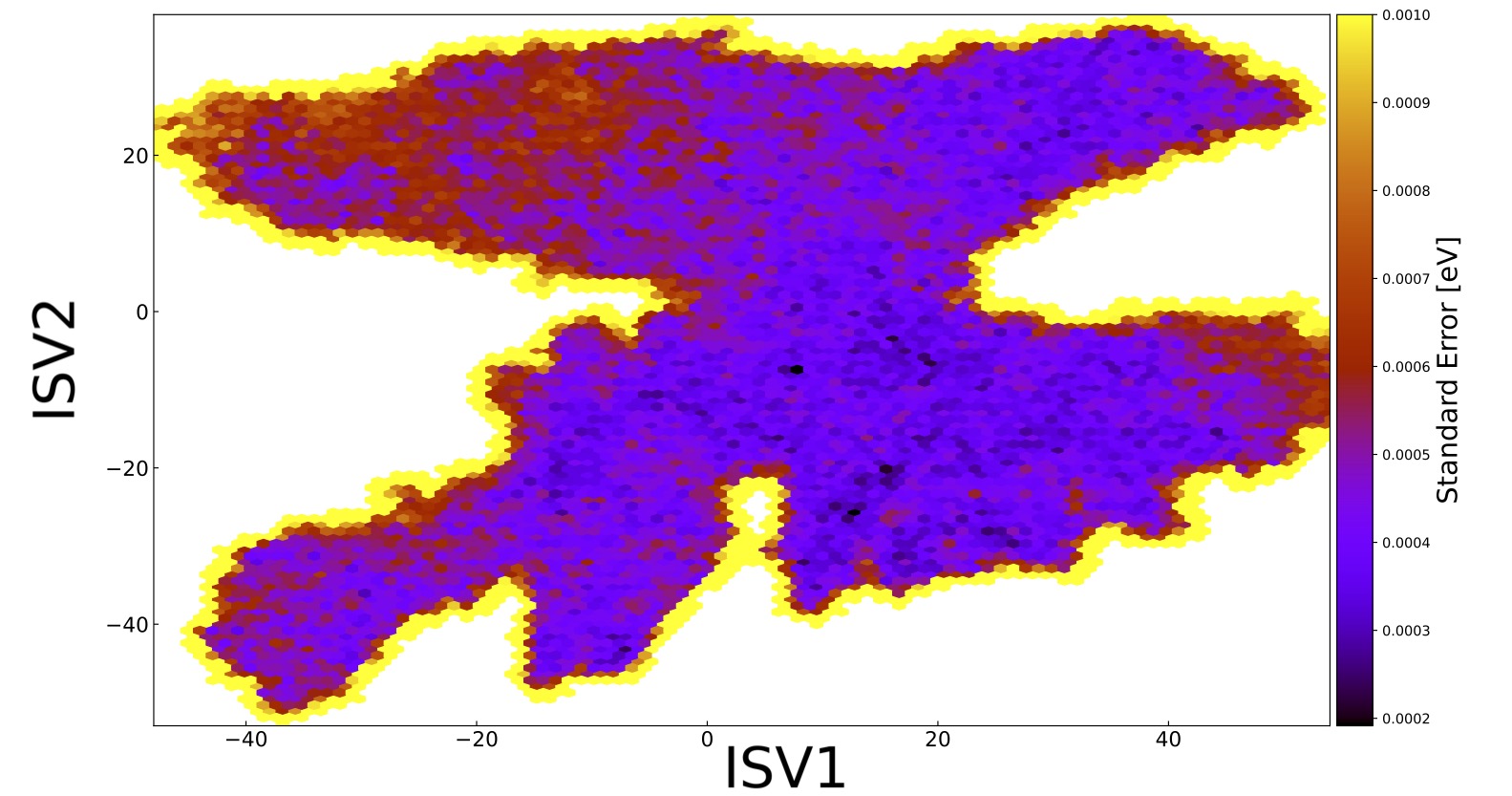}
        \caption[]{\textbf{Standard error of WHAM calculations.} 
        Plot of the standard error obtained via bootstrap calculations of the Free Energy computed via WHAM, reported in Fig.~2 of the main text.}
        \label{fig:stdboot}
\end{figure}

Umbrella sampling simulations were performed using a custom Metropolis Monte Carlo code written in C++ (see Supplementary Information, Sec. Monte Carlo Code). A total of 15.022 simulations, distributed over the range of explored ISVs values (Fig.~\ref{fig:us_points}) were performed. In each simulation, we took as starting coordinates the thermalized structure in the training dataset that was closer to the reference ISV value. 
The simulations consisted in a total of $20\cdot10^{6}$ MC moves, with the ISV restrained by a harmonic potential and a spring constant of 0.1 eV, which ensured a proper overlap of distributions between neighbor windows. MC moves were single atom gaussian displacement with a standard deviation of 0.07 nm. ISVs values were sampled every $5\cdot10^{3}$ moves for a total of $4\cdot10^{3}$ samples for every simulation.
After discarding the first $1/4$ of total samples, random sampling with replacement was used to generate 10 different data sets from the original population, selecting a number of samples equal to the half of the total, to account for time correlations. 
The total Free Energy (FE) landscape was then reconstructed with the  WHAM algorithm implemented by Grossfield\cite{grosswham}. The statistical error was estimated via the boostrapping method\cite{bootstrap} (Fig.~\ref{fig:stdboot}).

\FloatBarrier
\subsection*{Monte Carlo Code}
\label{si:mc}
The MC code was written \emph{de novo}  in C++. The code implements the Gupta potential and attempts only one type of MC move: the single atom displacement. Energy is updated after the move. The energy updating happens by recalculating only the terms pertaining to the the 146 distances that are modified in every trial move.
The atom to displace is randomly selected at every move and the entity of the displacement is selected following a gaussian distribution with fixed variance. The gaussian displacement was preferred with respect to the uniform displacement as it was observed to offer a faster decorrelation and a quicker sampling. The variance of the gaussian displacement was tuned in order to achieve an acceptance ratio of about 0.5

The AE is loaded into the code using \texttt{libtorch}, and it is able to import a JIT-compiled version of the inference model. Since the most computationally demanding part of every iteration was related to the KDE estimation of the RDFs, we proceeded to optimize these operations only by re-computing the gaussian kernels associated to the 146 distances that are modified at every attempt. A biasing scheme was introduced making it possible to run umbrella simulations. This was simply obtained by adding the harmonic restraint to the energy that is evaluated for the acceptance criterion of the Metropolis algorithm.

The code was tested by thoroughly comparing its output with LAMMPS unbiased simulation runs for several test-cases.

\subsubsection*{Test with $\text{Au}_{7}$}
\begin{figure}
        \centering
        \includegraphics[width=10cm]{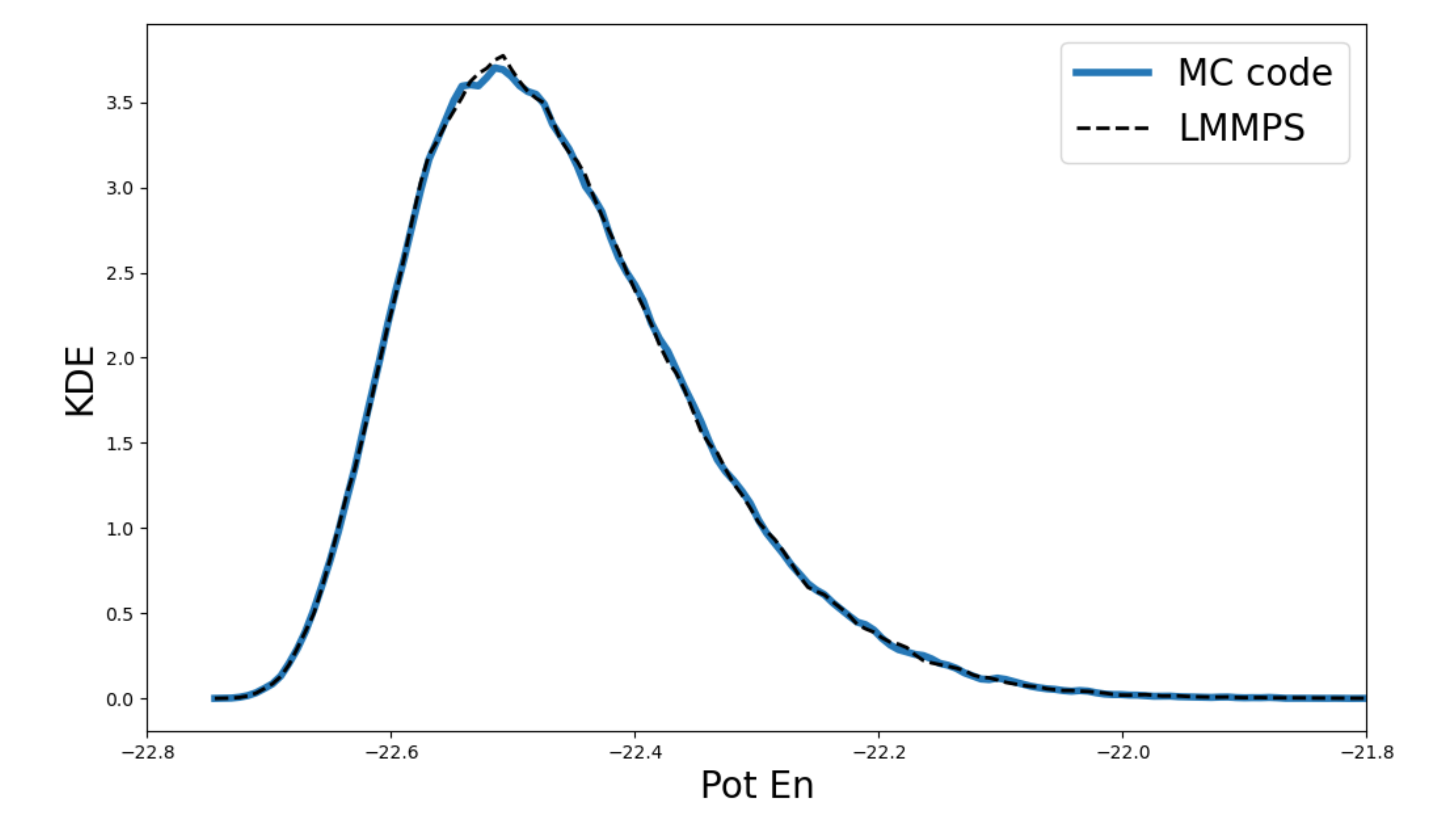}
        \caption[MC code sampling compared to LAMMPS for $\text{Au}_{7}$]{\textbf{MC comparison with LAMMPS for Au$_{7}$.} 
        Potential energy distributions for $\text{Au}_{7}$ sampled with the MC code (blue line) and LAMMPS (dashed black line).}
        \label{fig:potenau7}
\end{figure}

Firstly, a comparison using a smaller system was made to test the validity of the monte carlo scheme. In particular, we used a system composed of 7 gold atoms. The distribution of the potential energy obtained with LAMMPS was compared with the one obtained with our code.
Both simulations run for 1 billion iterations, printing every 5 thousand and discarding the first 50 thousand. Results are shown in Fig.~\ref{fig:potenau7}.

\subsubsection*{Test with $\text{Au}_{147}$}
\begin{figure}
        \centering
        \includegraphics[width=10cm]{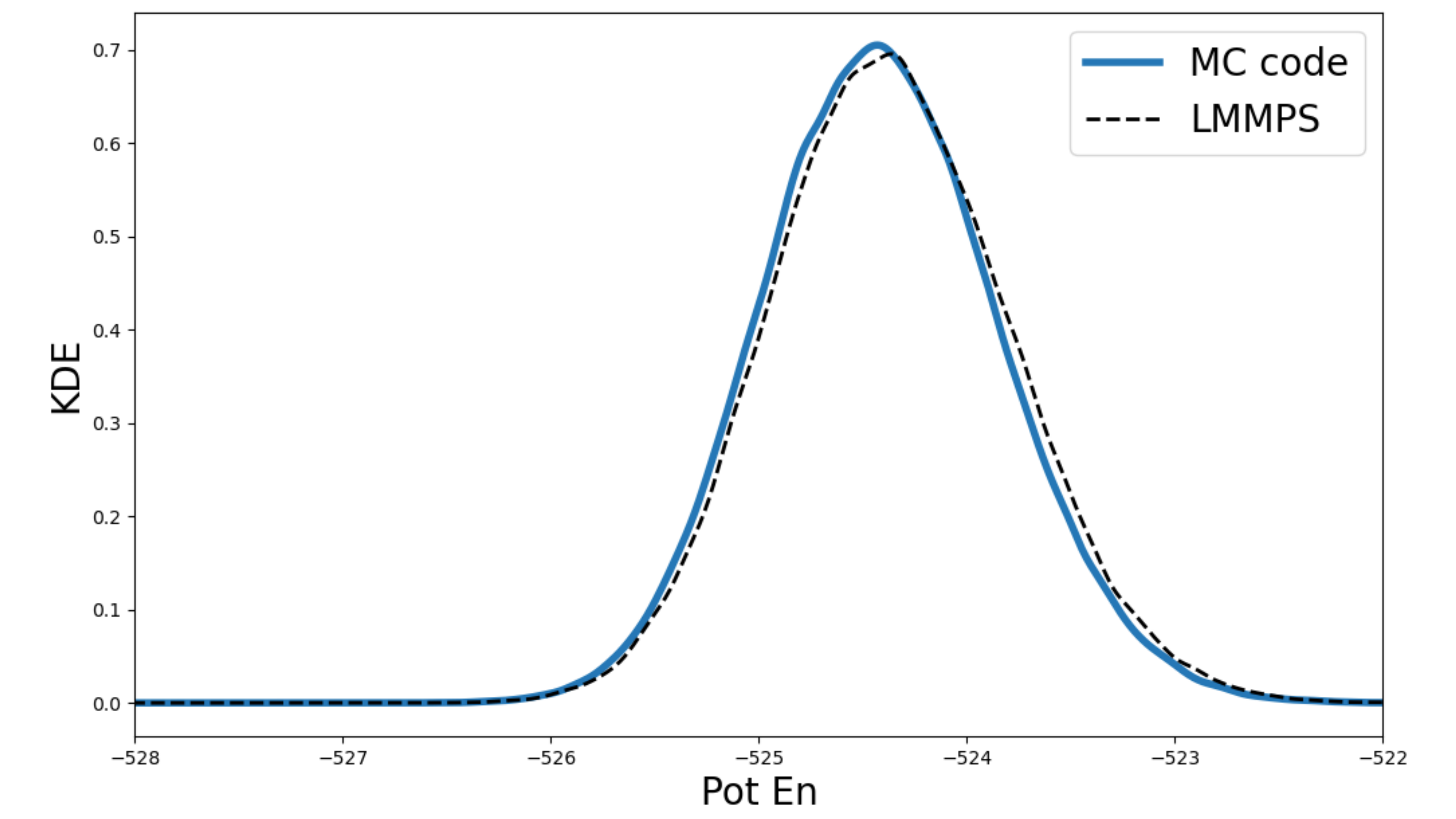}
        \caption{\textbf{MC comparison with LAMMPS for Au$_{147}$.}
        Potential energy distributions for $\text{Au}_{147}$ sampled with the MC code (blue line) and LAMMPS (dashed black line).}
        \label{fig:potenau147}
\end{figure}

After a similar test was conducted on $\text{Au}_{147}$. Again a LAMMPS simulation and an MC simulation were run starting from the same configuration, a fcc structure, running for 1 billion iterations, printing every 5 thousand and discarding the initial 50 thousand. Results are shown in Fig.~\ref{fig:potenau147}.

\subsubsection*{Sampling efficiency}
The MC code allowed for much faster iterations, with respect to LAMMPS code, in both unbiased and biased schemes (table~\ref{table:2}). This increased efficiency is mostly due to the limited number of calculations needed to update the energy and KDE calculations, thanks to single atom displacement.

\begin{table*}
\centering
\begin{tabular}{||l|c|c||}
\hline
 & MC code & LAMMPS\\
\hline\hline
Unbiased & 854603 & 384496\\
\hline
Biased & 179.5 & 8.4\\
\hline
\end{tabular}
\caption{\textbf{MC iteration per second comparison with LMMPS.} 
Table reporting the comparison between the number of iteration per second of the MC code versus LAMMPS code. Biasing in LAMMPS was applied wrapping the LAMMPS code with python to use the AE ISVs, in MC the ISVs were imported using libtorch. Benchmarks were conduced on a local machine, using a single core for all the simulations and averaging the performance over 1 million iterations}
\label{table:2}
\end{table*}

\begin{figure}
        \centering
        \includegraphics[width=\textwidth]{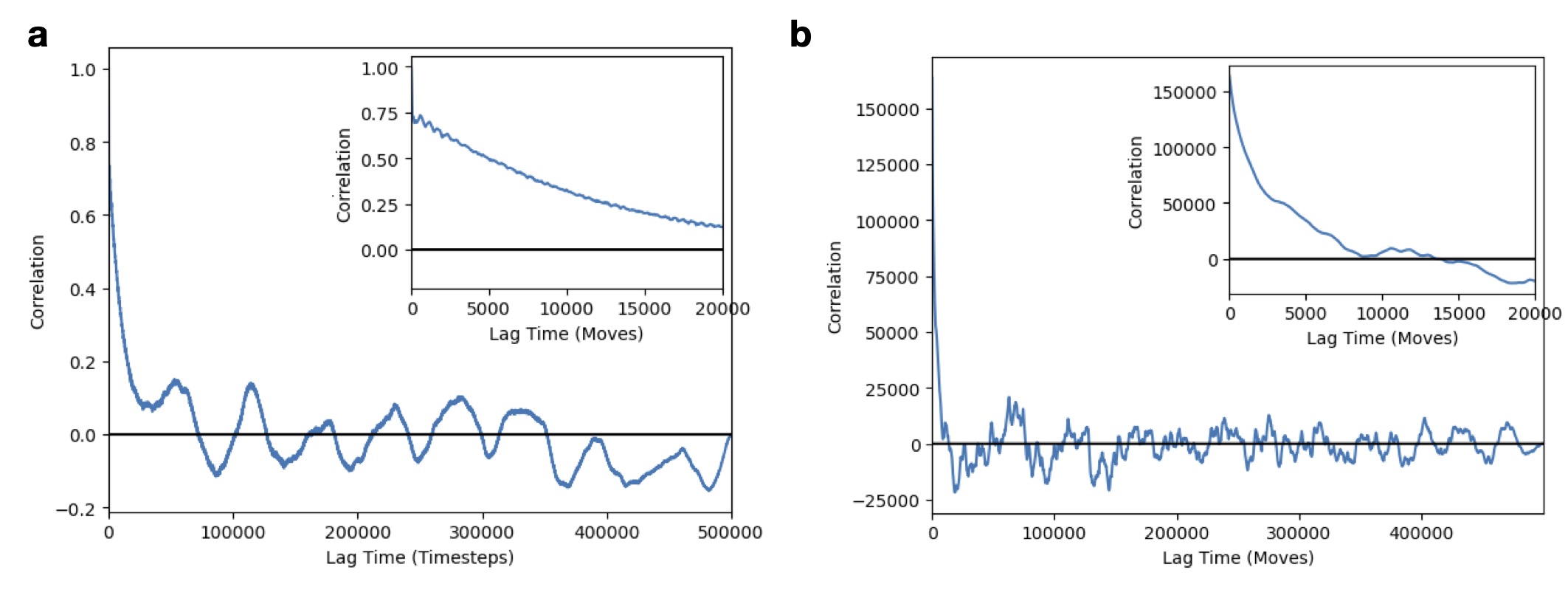}
        \caption{\textbf{MC autocorrelation comparison with MD.}
        \textbf{a} Autocorrelation of the potential energies of the sampled configurations in a LAMMPS unbiased simulation. \textbf{b} Autocorrelation of the potential energies of the sampled configurations in a MC simulation.}
        \label{fig:correlations}
\end{figure}

In order to compare the two different sampling schemes, and estimate if an actual increase in the generation of statistically valid samples was achieved, the autocorrelations of the generated data was considered, looking specifically at the autocorrelation of the potential energy. Results, shown in Fig.~\ref{fig:correlations}, proved that MC code allowed us for a much faster sampling.

\FloatBarrier

\subsection*{Markov State Models}

\subsubsection*{MSM simulations initialization}
\begin{figure}
        \centering
        \includegraphics[width=12cm]{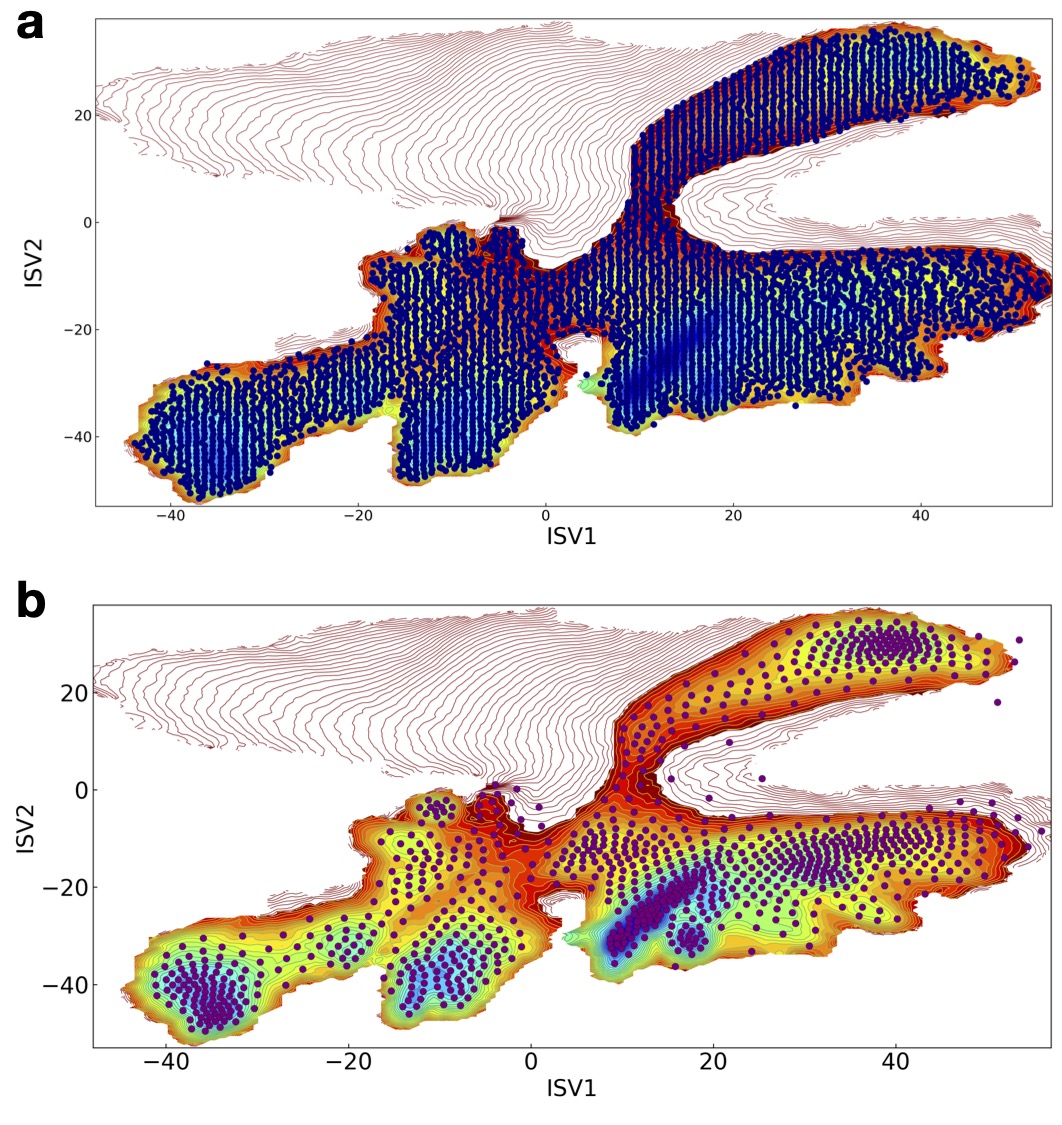}
        \caption{\textbf{Visualization of the initial configuration for the MD simulations and of the MSM and space discretization.}
        \textbf{a} Starting points of the 4448 MD simulations used for MSM. Every point is the initial configuration for a MD run. The point are plotted above the FE landscape of Fig.~2 of the main text.
        \textbf{b} Centers of the 1000 clusters obtained performing K-Means over the simulations outputs. They correspond to the 1000 states in which the ISVs space was discretized.}
        \label{fig:msmstartstate}
\end{figure}

In Fig.~\ref{fig:msmstartstate}a are reported the point in the ISV space where the simulations have been launched to gather trajectories for MSM analysis. In total 4448 simulations have been initialized all over the ISV space with lower FE values ($< 13 k_{b}T$).

\subsubsection*{ISV space discretization}
Trajectories were discretized in states, using K-Means over the samples ISVs, discretizing the space in 1000 clusters (Fig.~\ref{fig:msmstartstate}b).
\FloatBarrier
\subsubsection*{Implied Time scales and lag-time}
\begin{figure}
        \centering
        \includegraphics[width=7cm]{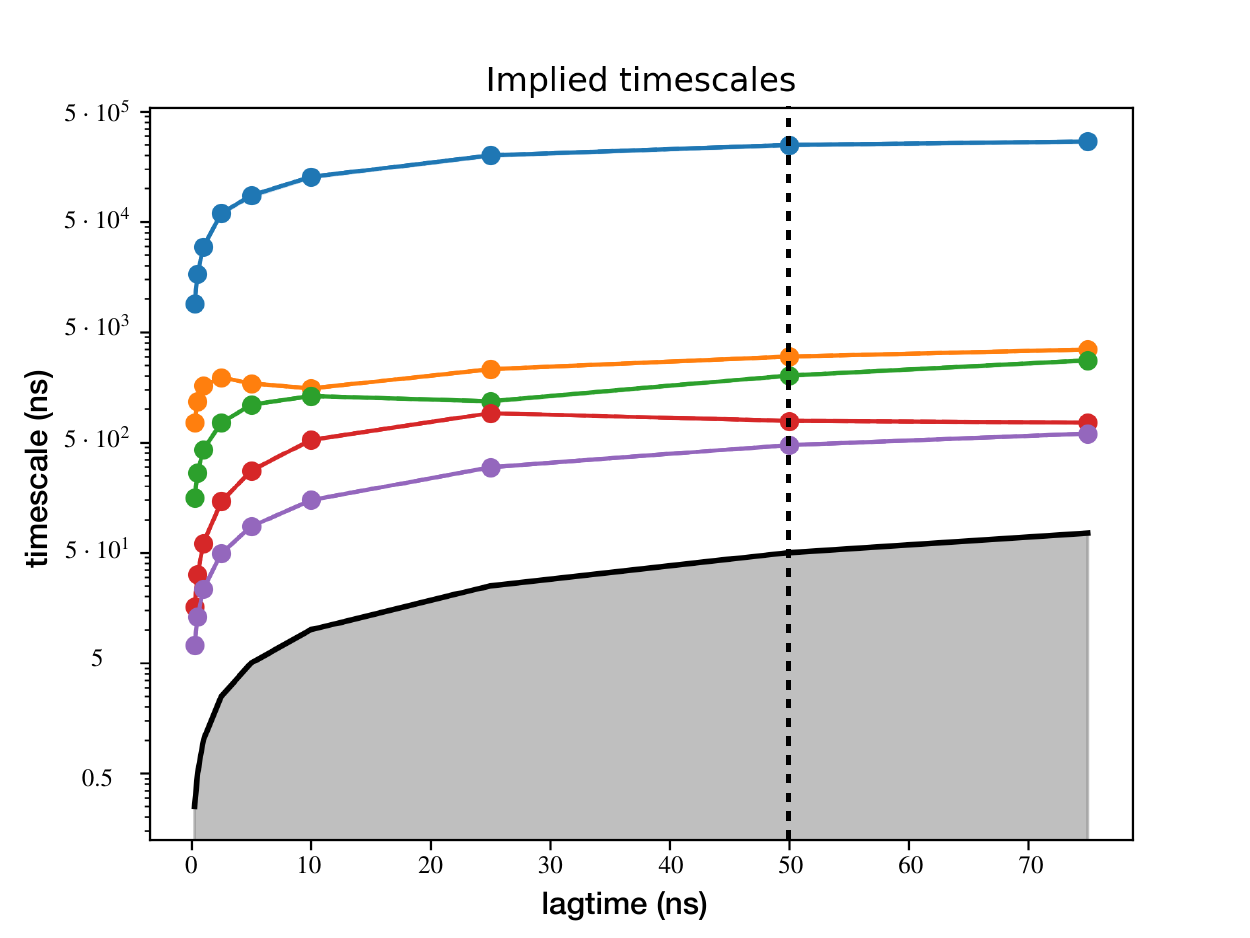}
        \caption{\textbf{Timescales convergence.}
        Implied time scales of the models as function of the lagtime. Based on the timescales convergence, the chosen lagtime was 50 ns. Analysis and plot obtained via DeepTime \cite{hoffmann2021deeptime}}
        \label{fig:its}
\end{figure}

The MSM lagtime was selected looking at the timescales convergence as illustrated in Fig.~\ref{fig:its}.
\FloatBarrier
\subsubsection*{Chapman-Kolmogorov test}
\begin{figure}[h!]
        \centering
        \includegraphics[width=8cm]{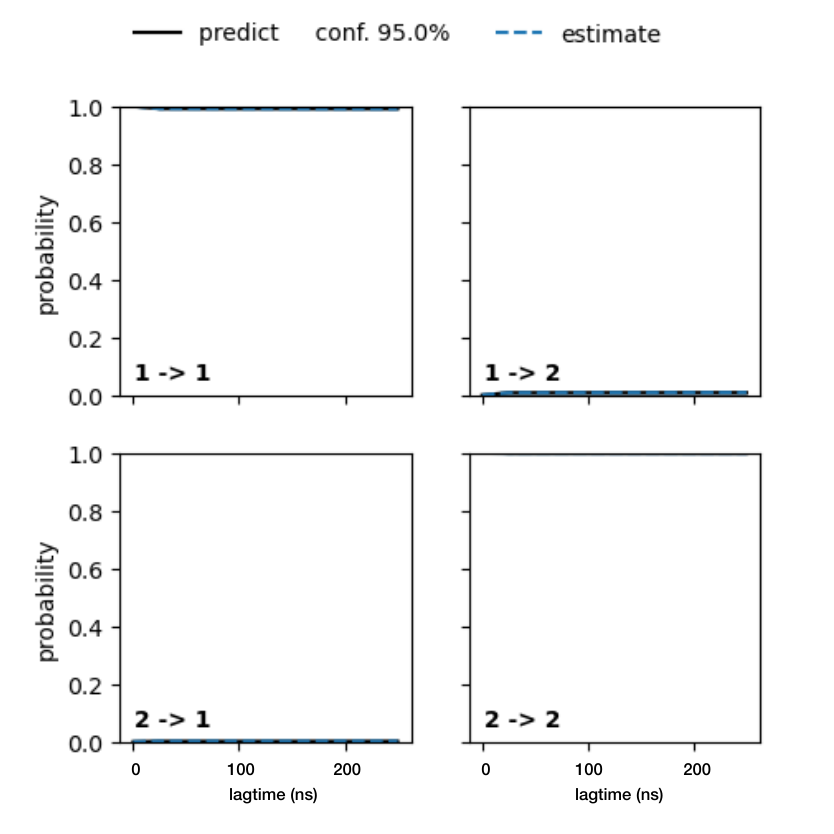}
        \caption{\textbf{2 states Chapman-Kolmogorov test.}
        Chapman-Kolmogorov test performed on the msm estimated model with 2 states. Plot obtained via DeepTime \cite{hoffmann2021deeptime}}
        \label{fig:ck2}
\end{figure}

\begin{figure}[h!]
        \centering
        \includegraphics[width=10cm]{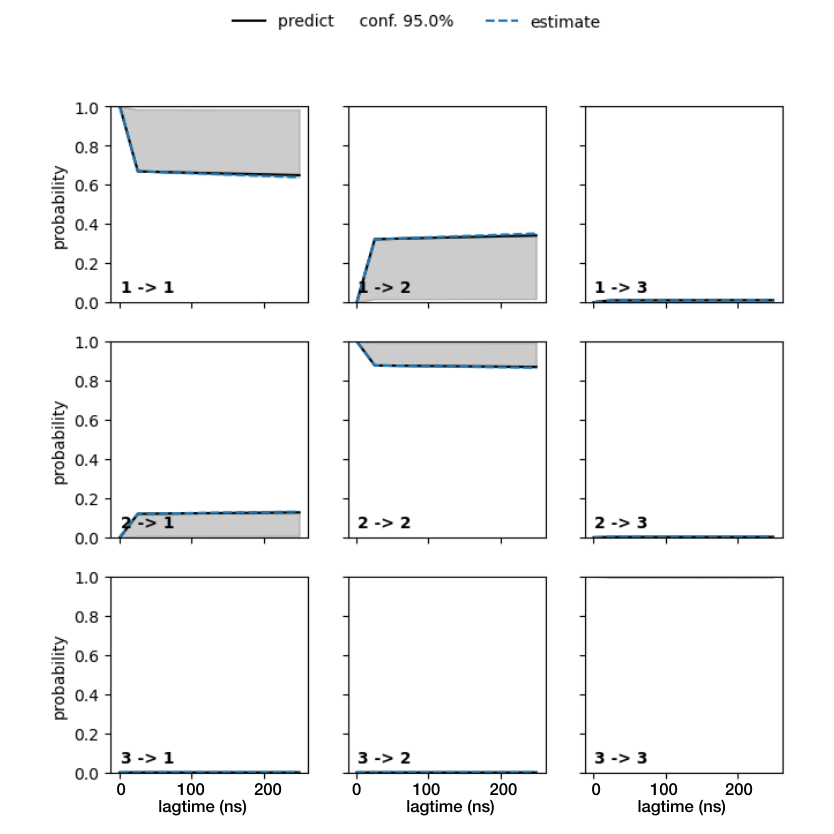}
        \caption{\textbf{3 states Chapman-Kolmogorov test.}
        Chapman-Kolmogorov test performed on the msm estimated model with 3 states. Plot obtained via DeepTime \cite{hoffmann2021deeptime}}
        \label{fig:ck3}
\end{figure}

To verify the quality of the markovian approximation, the Chapman-Kolmogorov test was performed using DeepTime library \cite{hoffmann2021deeptime}. Results are shown in Fig.~\ref{fig:ck2} and Fig.~\ref{fig:ck3}.
\paragraph{2 states Chapman-Kolmogorov test}
\FloatBarrier
\paragraph{3 states Chapman-Kolmogorov test}
\FloatBarrier
\subsubsection*{Committor Analysis from TPT}
\begin{figure}
        \centering
        \includegraphics[width= 9cm]{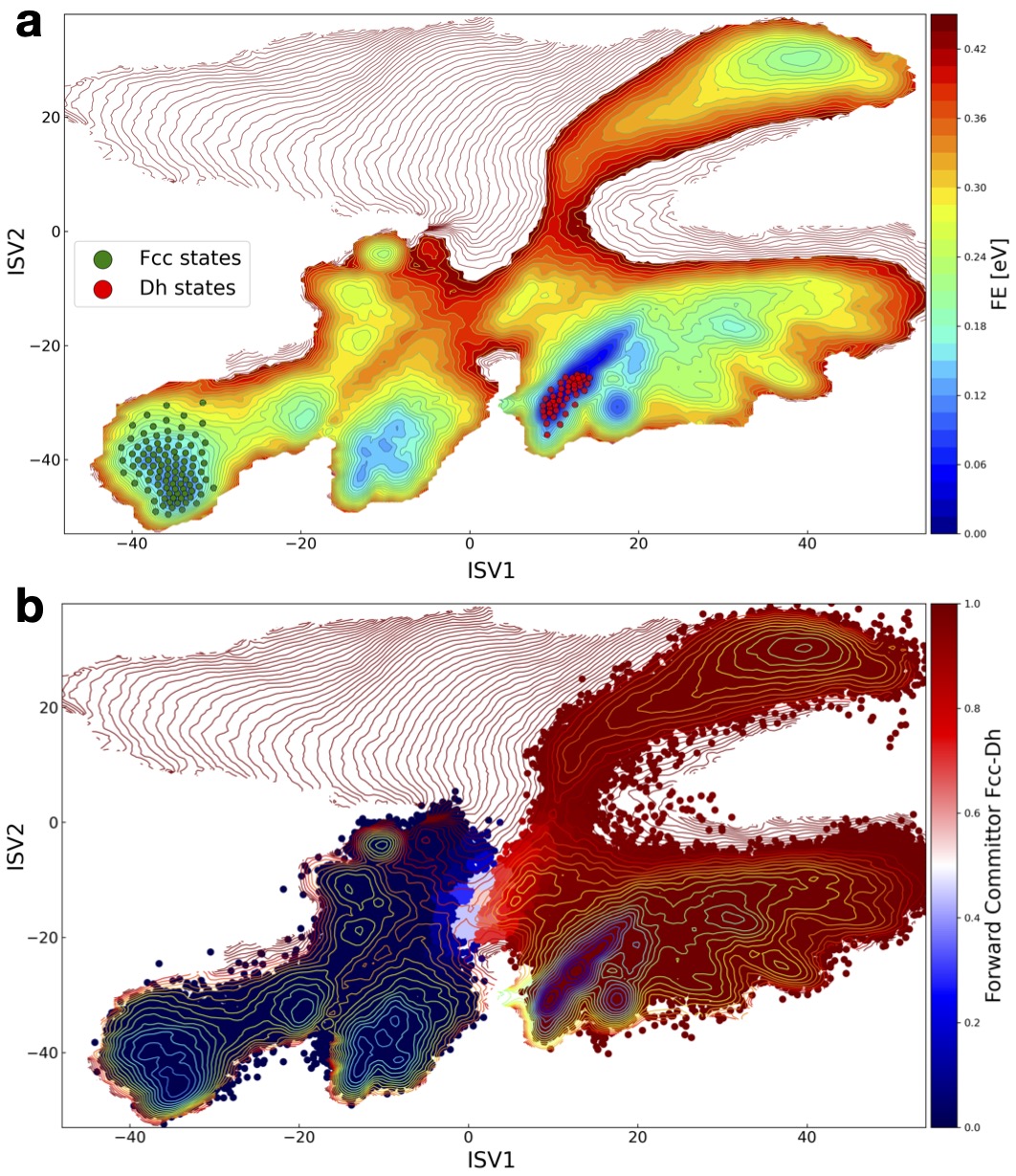}
        \caption{\textbf{Fcc-Dh transition committor calculations.}
        \textbf{a} Fcc (green points) and Dh (red points) basins definition for committor calculations. 
        \textbf{b} Forward committor plot for Fcc-Dh transition over the ISV space. The commitor represents the probability of a trajectory initialized in a specific point of the space to visit before one of the two states of the transition. The forward committor represents the probability to visit the final state before, in this case Dh. Same plot of Fig.~4a of the main text.}
        \label{fig:fccdhcomm}
\end{figure}

\begin{figure}
        \centering
        \includegraphics[width= 9cm]{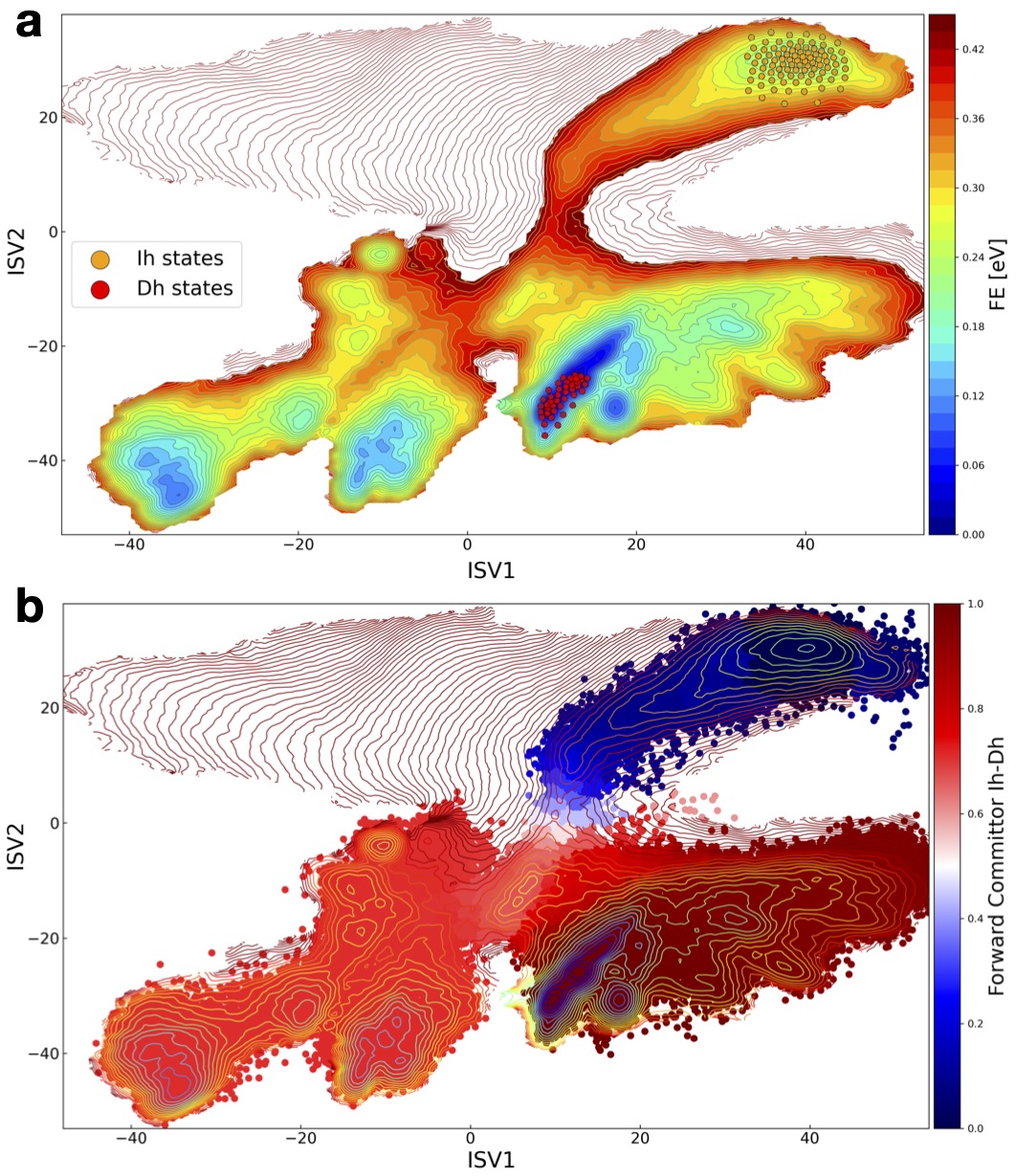}
        \caption{\textbf{Ih-Dh transition committor calculations.}
        \textbf{a} Ih (orange points) and Dh (red points) basins definition for committor calculations. 
        \textbf{b} Forward committor plot for Ih-Dh transition over the ISV space. The commitor represents the probability of a trajectory initialized in a specific point of the space to visit before one of the two states of the transition. The forward committor represents the probability to visit the final state before, in this case Dh.}
        \label{fig:ihdhcomm}
\end{figure}

Using the MSM and TPT analysis it was possible to estimate the committor value over the ISVs space for two specific transitions, namely Fcc to Dh and Ih to Dh. In order to compute the committor $q$, for each transitions were defined the two regions of space corresponding to initial and final state. In Fig.~\ref{fig:fccdhcomm}a we report the states among the 1000 in which the trajectories have been discretized (Fig.~\ref{fig:msmstartstate}b) selected to represent the Fcc basin (green points) and Dh basin (red points), while in Fig.~\ref{fig:fccdhcomm}b we report the estimated committor values for the different regions of the landscape. Similarly, in Fig.~\ref{fig:ihdhcomm}a the ones representing Ih basin (orange points) are reported, while the Dh basin definition was left untouched. The resulting commitor values for Ih-Dh transition is then reported in Fig.~\ref{fig:ihdhcomm}b.

\FloatBarrier


\begin{thebibliography}{66}
\providecommand{\natexlab}[1]{#1}
\providecommand{\url}[1]{\texttt{#1}}
\expandafter\ifx\csname urlstyle\endcsname\relax
  \providecommand{\doi}[1]{doi: #1}\else
  \providecommand{\doi}{doi: \begingroup \urlstyle{rm}\Url}\fi

\bibitem[Bolhuis et~al.(2002)Bolhuis, Chandler, Dellago, and Geissler]{bolhuis2002}
Peter~G Bolhuis, David Chandler, Christoph Dellago, and Phillip~L Geissler.
\newblock Transition path sampling: Throwing ropes over rough mountain passes, in the dark.
\newblock \emph{Annual review of physical chemistry}, 53\penalty0 (1):\penalty0 291--318, 2002.

\bibitem[Husic and Pande(2018)]{husic2018markov}
Brooke~E Husic and Vijay~S Pande.
\newblock Markov state models: From an art to a science.
\newblock \emph{Journal of the American Chemical Society}, 140\penalty0 (7):\penalty0 2386--2396, 2018.

\bibitem[Sprik and Ciccotti(1998)]{sprik1998free}
Michiel Sprik and Giovanni Ciccotti.
\newblock Free energy from constrained molecular dynamics.
\newblock \emph{The Journal of chemical physics}, 109\penalty0 (18):\penalty0 7737--7744, 1998.

\bibitem[Laio and Parrinello(2002)]{laio2002escaping}
Alessandro Laio and Michele Parrinello.
\newblock Escaping free-energy minima.
\newblock \emph{Proceedings of the national academy of sciences}, 99\penalty0 (20):\penalty0 12562--12566, 2002.

\bibitem[Stillinger and Weber(1983)]{stillinger1983inherent}
Frank~H Stillinger and Thomas~A Weber.
\newblock Inherent structure in water.
\newblock \emph{The Journal of Physical Chemistry}, 87\penalty0 (15):\penalty0 2833--2840, 1983.

\bibitem[Wales et~al.(1998)Wales, Miller, and Walsh]{wales1998archetypal}
David~J Wales, Mark~A Miller, and Tiffany~R Walsh.
\newblock Archetypal energy landscapes.
\newblock \emph{Nature}, 394\penalty0 (6695):\penalty0 758--760, 1998.

\bibitem[Baletto and Ferrando(2005)]{baletto2005rev}
F.~Baletto and R.~Ferrando.
\newblock Structural properties of nanoclusters: Energetic, thermodynamic, and kinetic effects.
\newblock \emph{Rev. Mod. Phys.}, 77:\penalty0 371--423, 2005.
\newblock \doi{10.1103/RevModPhys.77.371}.

\bibitem[Telari et~al.(2023)Telari, Tinti, Settem, Maragliano, Ferrando, and Giacomello]{telari2023}
E.~Telari, A.~Tinti, M.~Settem, L.~Maragliano, R.~Ferrando, and A.~Giacomello.
\newblock Charting nanocluster structures via convolutional neural networks.
\newblock \emph{ACS nano}, 17\penalty0 (21):\penalty0 21287--21296, 2023.

\bibitem[Sciortino et~al.(1999)Sciortino, Kob, and Tartaglia]{sciortino1999inherent}
Francesco Sciortino, W~Kob, and Piero Tartaglia.
\newblock Inherent structure entropy of supercooled liquids.
\newblock \emph{Physical Review Letters}, 83\penalty0 (16):\penalty0 3214, 1999.

\bibitem[Tanaka et~al.(2019)Tanaka, Tong, Shi, and Russo]{tanaka2019revealing}
Hajime Tanaka, Hua Tong, Rui Shi, and John Russo.
\newblock Revealing key structural features hidden in liquids and glasses.
\newblock \emph{Nature Reviews Physics}, 1\penalty0 (5):\penalty0 333--348, 2019.

\bibitem[Nakagawa and Peyrard(2006)]{nakagawa2006inherent}
Naoko Nakagawa and Michel Peyrard.
\newblock The inherent structure landscape of a protein.
\newblock \emph{Proceedings of the National Academy of Sciences}, 103\penalty0 (14):\penalty0 5279--5284, 2006.

\bibitem[Bolhuis et~al.(2000)Bolhuis, Dellago, and Chandler]{bolhuis2000reaction}
Peter~G Bolhuis, Christoph Dellago, and David Chandler.
\newblock Reaction coordinates of biomolecular isomerization.
\newblock \emph{Proceedings of the National Academy of Sciences}, 97\penalty0 (11):\penalty0 5877--5882, 2000.

\bibitem[Frenkel and Ladd(1984)]{frenkel1984new}
Daan Frenkel and Anthony~JC Ladd.
\newblock New monte carlo method to compute the free energy of arbitrary solids. application to the fcc and hcp phases of hard spheres.
\newblock \emph{The Journal of chemical physics}, 81\penalty0 (7):\penalty0 3188--3193, 1984.

\bibitem[Maragliano and Vanden-Eijnden(2006)]{maragliano2006temperature}
Luca Maragliano and Eric Vanden-Eijnden.
\newblock A temperature accelerated method for sampling free energy and determining reaction pathways in rare events simulations.
\newblock \emph{Chemical physics letters}, 426\penalty0 (1-3):\penalty0 168--175, 2006.

\bibitem[Allen et~al.(2009)Allen, Valeriani, and Ten~Wolde]{allen2009forward}
Rosalind~J Allen, Chantal Valeriani, and Pieter~Rein Ten~Wolde.
\newblock Forward flux sampling for rare event simulations.
\newblock \emph{Journal of physics: Condensed matter}, 21\penalty0 (46):\penalty0 463102, 2009.

\bibitem[Sultan and Pande(2018)]{sultan2018automated}
Mohammad~M Sultan and Vijay~S Pande.
\newblock Automated design of collective variables using supervised machine learning.
\newblock \emph{The Journal of chemical physics}, 149\penalty0 (9), 2018.

\bibitem[Wehmeyer and No{\'e}(2018)]{wehmeyer2018time}
Christoph Wehmeyer and Frank No{\'e}.
\newblock Time-lagged autoencoders: Deep learning of slow collective variables for molecular kinetics.
\newblock \emph{The Journal of chemical physics}, 148\penalty0 (24), 2018.

\bibitem[Sidky et~al.(2020)Sidky, Chen, and Ferguson]{sidky2020machine}
Hythem Sidky, Wei Chen, and Andrew~L Ferguson.
\newblock Machine learning for collective variable discovery and enhanced sampling in biomolecular simulation.
\newblock \emph{Molecular Physics}, 118\penalty0 (5):\penalty0 e1737742, 2020.

\bibitem[Jung et~al.(2023)Jung, Covino, Arjun, Leitold, Dellago, Bolhuis, and Hummer]{jung2023machine}
Hendrik Jung, Roberto Covino, A~Arjun, Christian Leitold, Christoph Dellago, Peter~G Bolhuis, and Gerhard Hummer.
\newblock Machine-guided path sampling to discover mechanisms of molecular self-organization.
\newblock \emph{Nature Computational Science}, 3\penalty0 (4):\penalty0 334--345, 2023.

\bibitem[Pavan et~al.(2015)Pavan, Rossi, and Baletto]{pavan2015NPsMeetMetaD}
L.~Pavan, K.~Rossi, and F.~Baletto.
\newblock Metallic nanoparticles meet metadynamics.
\newblock \emph{J. Chem. Phys.}, 143:\penalty0 184304, 2015.
\newblock \doi{10.1063/1.4935272}.

\bibitem[Tribello et~al.(2017)Tribello, Giberti, Sosso, Salvalaglio, and Parrinello]{tribello2017}
Gareth~A Tribello, Federico Giberti, Gabriele~C Sosso, Matteo Salvalaglio, and Michele Parrinello.
\newblock Analyzing and driving cluster formation in atomistic simulations.
\newblock \emph{Journal of chemical theory and computation}, 13\penalty0 (3):\penalty0 1317--1327, 2017.

\bibitem[Pipolo et~al.(2017)Pipolo, Salanne, Ferlat, Klotz, Saitta, and Pietrucci]{pipolo2017navigating}
Silvio Pipolo, Mathieu Salanne, Guillaume Ferlat, Stefan Klotz, A~Marco Saitta, and Fabio Pietrucci.
\newblock Navigating at will on the water phase diagram.
\newblock \emph{Physical review letters}, 119\penalty0 (24):\penalty0 245701, 2017.

\bibitem[Fan et~al.(2017)Fan, Iwashita, and Egami]{fan2017energy}
Yue Fan, Takuya Iwashita, and Takeshi Egami.
\newblock Energy landscape-driven non-equilibrium evolution of inherent structure in disordered material.
\newblock \emph{Nature communications}, 8\penalty0 (1):\penalty0 15417, 2017.

\bibitem[Rao and Karplus(2010)]{rao2010protein}
Francesco Rao and Martin Karplus.
\newblock Protein dynamics investigated by inherent structure analysis.
\newblock \emph{Proceedings of the National Academy of Sciences}, 107\penalty0 (20):\penalty0 9152--9157, 2010.

\bibitem[Foster et~al.(2018)Foster, Ferrando, and Palmer]{foster2018AuImagingACFraction}
D.~M. Foster, R.~Ferrando, and R.~E. Palmer.
\newblock Experimental determination of the energy difference between competing isomers of deposited, size-selected gold nanoclusters.
\newblock \emph{Nat. Commun.}, 9:\penalty0 1323, 2018.
\newblock \doi{10.1038/s41467-018-03794-9}.

\bibitem[Mottet et~al.(1997)Mottet, Tr\'eglia, and Legrand]{Mottet1997ss}
C.~Mottet, G.~Tr\'eglia, and B.~Legrand.
\newblock New magic numbers in metallic clusters: an unexpected metal dependence.
\newblock \emph{Surf. Sci.}, 383:\penalty0 L719--L727, 1997.

\bibitem[Apra et~al.(2004)Apra, Baletto, Ferrando, and Fortunelli]{apra2004AuRosette}
E.~Apra, F.~Baletto, R.~Ferrando, and A.~Fortunelli.
\newblock Amorphization mechanism of icosahedral metal nanoclusters.
\newblock \emph{Phys. Rev. Lett.}, 93:\penalty0 065502, 2004.
\newblock \doi{10.1103/PhysRevLett.93.065502}.

\bibitem[Xia et~al.(2021)Xia, Nelli, Ferrando, Yuan, and Li]{Xia2021ncomms}
Yu~Xia, Diana Nelli, Riccardo Ferrando, Jun Yuan, and Z.~Y. Li.
\newblock Shape control of size-selected naked platinum nanocrystals.
\newblock \emph{Nature Communications}, 12:\penalty0 3019, 2021.

\bibitem[Amodeo et~al.(2020)Amodeo, Pietrucci, and Lam]{amodeo2020}
Jonathan Amodeo, Fabio Pietrucci, and Julien Lam.
\newblock Out-of-equilibrium polymorph selection in nanoparticle freezing.
\newblock \emph{The Journal of Physical Chemistry Letters}, 11\penalty0 (19):\penalty0 8060--8066, 2020.

\bibitem[Settem et~al.(2022)Settem, Ferrando, and Giacomello]{settem2022AuPTMD}
M.~Settem, R.~Ferrando, and A.~Giacomello.
\newblock Tempering of {Au} nanoclusters: capturing the temperature-dependent competition among structural motifs.
\newblock \emph{Nanoscale}, 14:\penalty0 939--952, 2022.
\newblock \doi{10.1039/D1NR05078H}.

\bibitem[Settem et~al.(2023)Settem, Roncaglia, Ferrando, and Giacomello]{settemAgCu}
Manoj Settem, Cesare Roncaglia, Riccardo Ferrando, and Alberto Giacomello.
\newblock Structural transformations in \uppercase{C}u, \uppercase{A}g, and \uppercase{A}u metal nanoclusters.
\newblock \emph{The Journal of Chemical Physics}, 159\penalty0 (9), 2023.

\bibitem[Schebarchov et~al.(2018)Schebarchov, Baletto, and Wales]{schebarchov2018AuHSA}
D.~Schebarchov, F.~Baletto, and D.~J. Wales.
\newblock Structure, thermodynamics, and rearrangement mechanisms in gold clusters—insights from the energy landscapes framework.
\newblock \emph{Nanoscale}, 10:\penalty0 2004--2016, 2018.
\newblock \doi{10.1039/c7nr07123j}.

\bibitem[Calvo et~al.(2002)Calvo, Doye, and Wales]{calvo2002}
F~Calvo, JPK Doye, and DJ~Wales.
\newblock Equilibrium properties of clusters in the harmonic superposition approximation.
\newblock \emph{Chemical physics letters}, 366\penalty0 (1-2):\penalty0 176--183, 2002.

\bibitem[Hinton and Salakhutdinov(2006)]{hinton2006autoencoder}
Geoffrey~E Hinton and Ruslan~R Salakhutdinov.
\newblock Reducing the dimensionality of data with neural networks.
\newblock \emph{Science}, 313\penalty0 (5786):\penalty0 504--507, 2006.

\bibitem[Rosato et~al.(1989)Rosato, Guillope, and Legrand]{rosato1989}
Vittorio Rosato, M~Guillope, and B~Legrand.
\newblock Thermodynamical and structural properties of fcc transition metals using a simple tight-binding model.
\newblock \emph{Philosophical Magazine A}, 59\penalty0 (2):\penalty0 321--336, 1989.

\bibitem[Santarossa et~al.(2010)Santarossa, Vargas, Iannuzzi, and Baiker]{santarossa2010free}
Gianluca Santarossa, Angelo Vargas, Marcella Iannuzzi, and Alfons Baiker.
\newblock Free energy surface of two-and three-dimensional transitions of au 12 nanoclusters obtained by ab initio metadynamics.
\newblock \emph{Physical Review B—Condensed Matter and Materials Physics}, 81\penalty0 (17):\penalty0 174205, 2010.

\bibitem[El~koraychy et~al.(2022)El~koraychy, Roncaglia, Nelli, Cerbelaud, and Ferrando]{Elkoraychy2022nh}
El~yakout El~koraychy, Cesare Roncaglia, Diana Nelli, Manuella Cerbelaud, and Riccardo Ferrando.
\newblock Growth mechanisms from tetrahedral seeds to multiply twinned au nanoparticles revealed by atomistic simulations.
\newblock \emph{Nanoscale Horiz.}, 7:\penalty0 883--889, 2022.
\newblock \doi{10.1039/D1NH00599E}.

\bibitem[Wang and Palmer(2012{\natexlab{a}})]{wang2012AuImaging}
Z.~W. Wang and R.~E. Palmer.
\newblock Determination of the ground-state atomic structures of size-selected au nanoclusters by electron-beam-induced transformation.
\newblock \emph{Phys. Rev. Lett.}, 108:\penalty0 245502, 2012{\natexlab{a}}.
\newblock \doi{10.1103/PhysRevLett.108.245502}.

\bibitem[Bowman et~al.(2013)Bowman, Pande, and No{\'e}]{bowman2013introduction}
Gregory~R Bowman, Vijay~S Pande, and Frank No{\'e}.
\newblock \emph{An introduction to Markov state models and their application to long timescale molecular simulation}, volume 797.
\newblock Springer Science \& Business Media, 2013.

\bibitem[Metzner et~al.(2009)Metzner, Sch{\"u}tte, and Vanden-Eijnden]{metznerTPT}
Philipp Metzner, Christof Sch{\"u}tte, and Eric Vanden-Eijnden.
\newblock Transition path theory for markov jump processes.
\newblock \emph{Multiscale Modeling \& Simulation}, 7\penalty0 (3):\penalty0 1192--1219, 2009.

\bibitem[E et~al.(2002)E, Ren, and Vanden-Eijnden]{e2002string}
Weinan E, Weiqing Ren, and Eric Vanden-Eijnden.
\newblock String method for the study of rare events.
\newblock \emph{Physical Review B}, 66\penalty0 (5):\penalty0 052301, 2002.

\bibitem[Huang et~al.(2018)Huang, Wen, Voter, and Perez]{Huang2018prm}
Rao Huang, Yuhua Wen, Arthur~F. Voter, and Danny Perez.
\newblock Direct observations of shape fluctuation in long-time atomistic simulations of metallic nanoclusters.
\newblock \emph{Phys. Rev. Materials}, 2:\penalty0 126002, 2018.
\newblock \doi{10.1103/PhysRevMaterials.2.126002}.

\bibitem[Dearg et~al.(2024{\natexlab{a}})Dearg, Roncaglia, Nelli, Ferrando, Slater, Palmer, et~al.]{dearg2024frame}
Malcolm Dearg, Cesare Roncaglia, Diana Nelli, Riccardo Ferrando, Thomas~JA Slater, Richard~E Palmer, et~al.
\newblock Frame-by-frame observations of structure fluctuations in single mass-selected au clusters using aberration-corrected electron microscopy.
\newblock \emph{Nanoscale Horizons}, 9\penalty0 (1):\penalty0 143--147, 2024{\natexlab{a}}.

\bibitem[Torrie and Valleau(1974)]{torrie1974monte}
Glenn~M Torrie and John~P Valleau.
\newblock Monte carlo free energy estimates using non-boltzmann sampling: Application to the sub-critical lennard-jones fluid.
\newblock \emph{Chemical Physics Letters}, 28\penalty0 (4):\penalty0 578--581, 1974.

\bibitem[De~Nijs et~al.(2015)De~Nijs, Dussi, Smallenburg, Meeldijk, Groenendijk, Filion, Imhof, Van~Blaaderen, and Dijkstra]{de2015entropy}
Bart De~Nijs, Simone Dussi, Frank Smallenburg, Johannes~D Meeldijk, Dirk~J Groenendijk, Laura Filion, Arnout Imhof, Alfons Van~Blaaderen, and Marjolein Dijkstra.
\newblock Entropy-driven formation of large icosahedral colloidal clusters by spherical confinement.
\newblock \emph{Nature materials}, 14\penalty0 (1):\penalty0 56--60, 2015.

\bibitem[Ferrenberg and Swendsen(1989)]{ferrenberg1989}
Alan~M Ferrenberg and Robert~H Swendsen.
\newblock Optimized monte carlo data analysis.
\newblock \emph{Physical review letters}, 63\penalty0 (12):\penalty0 1195, 1989.

\bibitem[Grossfield()]{grosswham}
Alan Grossfield.
\newblock Wham: the weighted histogram analysis method, version 2.0.11.
\newblock \doi{http://membrane.urmc.rochester.edu/wordpress/?page_id=126}.

\bibitem[Thompson et~al.(2022)Thompson, Aktulga, Berger, Bolintineanu, Brown, Crozier, in~'t Veld, Kohlmeyer, Moore, Nguyen, Shan, Stevens, Tranchida, Trott, and Plimpton]{LAMMPS}
A.~P. Thompson, H.~M. Aktulga, R.~Berger, D.~S. Bolintineanu, W.~M. Brown, P.~S. Crozier, P.~J. in~'t Veld, A.~Kohlmeyer, S.~G. Moore, T.~D. Nguyen, R.~Shan, M.~J. Stevens, J.~Tranchida, C.~Trott, and S.~J. Plimpton.
\newblock {LAMMPS} - a flexible simulation tool for particle-based materials modeling at the atomic, meso, and continuum scales.
\newblock \emph{Comp. Phys. Comm.}, 271:\penalty0 108171, 2022.
\newblock \doi{10.1016/j.cpc.2021.108171}.

\bibitem[Hoffmann et~al.(2021)Hoffmann, Scherer, Hempel, Mardt, de~Silva, Husic, Klus, Wu, Kutz, Brunton, et~al.]{hoffmann2021deeptime}
Moritz Hoffmann, Martin Scherer, Tim Hempel, Andreas Mardt, Brian de~Silva, Brooke~E Husic, Stefan Klus, Hao Wu, Nathan Kutz, Steven~L Brunton, et~al.
\newblock Deeptime: a python library for machine learning dynamical models from time series data.
\newblock \emph{Machine Learning: Science and Technology}, 3\penalty0 (1):\penalty0 015009, 2021.

\bibitem[No{\'e} et~al.(2009)No{\'e}, Sch{\"u}tte, Vanden-Eijnden, Reich, and Weikl]{noeCommittor}
Frank No{\'e}, Christof Sch{\"u}tte, Eric Vanden-Eijnden, Lothar Reich, and Thomas~R Weikl.
\newblock Constructing the equilibrium ensemble of folding pathways from short off-equilibrium simulations.
\newblock \emph{Proceedings of the National Academy of Sciences}, 106\penalty0 (45):\penalty0 19011--19016, 2009.

\bibitem[Wang and Palmer(2012{\natexlab{b}})]{wang2012Au20}
Z.~W. Wang and R.~E. Palmer.
\newblock Direct atomic imaging and dynamical fluctuations of the tetrahedral au20 cluster.
\newblock \emph{Nanoscale}, 4:\penalty0 4947--4949, 2012{\natexlab{b}}.
\newblock \doi{10.1039/C2NR31071F}.

\bibitem[Wang and Palmer(2012{\natexlab{c}})]{wang2012Au55}
Z.~W. Wang and R.~E. Palmer.
\newblock Experimental evidence for fluctuating, chiral-type au$_{55}$ clusters by direct atomic imaging.
\newblock \emph{Nano Lett.}, 12:\penalty0 5510--5514, 2012{\natexlab{c}}.
\newblock \doi{10.1021/nl303429z}.

\bibitem[Pratontep et~al.(2005)Pratontep, Carroll, Xirouchaki, Streun, and Palmer]{pratontep2005}
S.~Pratontep, S.~J. Carroll, C.~Xirouchaki, M.~Streun, and R.~E. Palmer.
\newblock Size-selected cluster beam source based on radio frequency magnetron plasma sputtering and gas condensation.
\newblock \emph{Rev. Sci. Instrum.}, 76:\penalty0 045103, 2005.
\newblock \doi{10.1063/1.1869332}.

\bibitem[Issendorff and Palmer(1999)]{issendorff1999}
B.~von Issendorff and R.~E. Palmer.
\newblock A new high transmission infinite range mass selector for cluster and nanoparticle beams.
\newblock \emph{Rev. Sci. Instrum.}, 70:\penalty0 4497--4501, 1999.
\newblock \doi{10.1063/1.1150102}.

\bibitem[Di~Vece et~al.(2005)Di~Vece, Palomba, and Palmer]{diVece2005}
M.~Di~Vece, S.~Palomba, and R.~E. Palmer.
\newblock Pinning of size-selected gold and nickel nanoclusters on graphite.
\newblock \emph{Phys. Rev. B}, 72:\penalty0 073407, 2005.
\newblock \doi{10.1103/PhysRevB.72.073407}.

\bibitem[Claeyssens et~al.(2006)Claeyssens, Pratontep, Xirouchaki, and Palmer]{claeyssens2006}
F.~Claeyssens, S.~Pratontep, C.~Xirouchaki, and R.~E. Palmer.
\newblock Immobilization of large size-selected silver clusters on graphite.
\newblock \emph{Nanotechnology}, 17:\penalty0 805--807, 2006.
\newblock \doi{10.1088/0957-4484/17/3/032}.

\bibitem[Madsen and Susi(2021)]{madsen2021abTEM}
J.~Madsen and T.~Susi.
\newblock The abtem code: transmission electron microscopy from first principles [version 2; peer review: 2 approved].
\newblock \emph{Open Research Europe}, 1:\penalty0 24, 2021.
\newblock \doi{10.12688/openreseurope.13015.2}.

\bibitem[Ophus(2017)]{ophus2017}
C.~Ophus.
\newblock A fast image simulation algorithm for scanning transmission electron microscopy.
\newblock \emph{Adv. Struct. Chem. Imaging}, 3:\penalty0 13, 2017.
\newblock \doi{10.1186/s40679-017-0046-1}.

\bibitem[Wang et~al.(2011)Wang, Li, Park, Abdela, Tang, and Palmer]{wang2011}
Z.~W. Wang, Z.~Y. Li, S.~J. Park, A.~Abdela, D.~Tang, and R.~E. Palmer.
\newblock Quantitative z-contrast imaging in the scanning transmission electron microscope with size-selected clusters.
\newblock \emph{Phys. Rev. B}, 84:\penalty0 073408, 2011.
\newblock \doi{10.1103/PhysRevB.84.073408}.

\bibitem[Dearg et~al.(2024{\natexlab{b}})Dearg, Lethbridge, McCormack, Palmer, and A.~Slater]{dearg2024}
M.~Dearg, S.~Lethbridge, J.~McCormack, R.~E. Palmer, and T.~J. A.~Slater.
\newblock Characterisation of the morphology of surface-assembled au nanoclusters on amorphous carbon.
\newblock \emph{Nanoscale}, 16:\penalty0 10827--10832, 2024{\natexlab{b}}.
\newblock \doi{10.1039/D4NR00978A}.

\bibitem[Lethbridge et~al.(2024)Lethbridge, Pavloudis, Slater, Kioseoglou, and Palmer]{lethbridge2024}
S.~Lethbridge, T.~M. Pavloudis, T.~J. Slater, J.~Kioseoglou, and R.~E. Palmer.
\newblock Stabilization of 2d raft structures of au nanoclusters with up to 60 atoms by a carbon support.
\newblock \emph{Small Sci.}, 2024.
\newblock \doi{10.1002/smsc.202400093}.

\bibitem[Rosenblatt(1956)]{rosenblattKDE}
Murray Rosenblatt.
\newblock A central limit theorem and a strong mixing condition.
\newblock \emph{Proceedings of the national Academy of Sciences}, 42\penalty0 (1):\penalty0 43--47, 1956.

\bibitem[Parzen(1962)]{parzenKDE}
Emanuel Parzen.
\newblock On estimation of a probability density function and mode.
\newblock \emph{The annals of mathematical statistics}, 33\penalty0 (3):\penalty0 1065--1076, 1962.

\bibitem[Paszke et~al.(2017)Paszke, Gross, Chintala, Chanan, Yang, DeVito, Lin, Desmaison, Antiga, and Lerer]{paszke2017automatic}
Adam Paszke, Sam Gross, Soumith Chintala, Gregory Chanan, Edward Yang, Zachary DeVito, Zeming Lin, Alban Desmaison, Luca Antiga, and Adam Lerer.
\newblock Automatic differentiation in pytorch.
\newblock 2017.

\bibitem[Kingma and Ba(2017)]{adam}
Diederik~P. Kingma and Jimmy Ba.
\newblock Adam: A method for stochastic optimization, 2017.

\bibitem[Efron(1992)]{bootstrap}
Bradley Efron.
\newblock \emph{Bootstrap methods: another look at the jackknife}, pages 569--593.
\newblock Springer, Berlin, 1992.

\end{thebibliography}
\end{document}